\documentclass[journal=jacsat,manuscript=article]{achemso}

\usepackage{physics}
\usepackage[T1]{fontenc}
\usepackage{multirow}
\usepackage{amsmath}

\author{Aleksandra Leszczyk}
\affiliation{Institute of Physics, Nicolaus Copernicus University in Toru\'{n}, Toru{n}, Poland}

\author{Mih\'{a}ly M\'{a}t\'{e}}
\affiliation{Strongly Correlated Systems "Lend\"{u}let" Research Group, Wigner Research Center for Physics, H-1525 Budapest, Hungary}

\author{\"{O}rs Legeza}
\affiliation{Strongly Correlated Systems "Lend\"{u}let" Research Group, Wigner Research Center for Physics, H-1525 Budapest, Hungary}

\author{Katharina Boguslawski}
\email{k.boguslawski@fizyka.umk.pl}
\affiliation{Institute of Physics, Nicolaus Copernicus University in Toru\'{n}, Toru{n}, Poland}

\title[]
  {Assessing the accuracy of tailored coupled cluster methods corrected by electronic wave functions of polynomial cost.}

\abbreviations{DMRG,pCCD,tCC}
\keywords{Coupled Cluster, DMRG, pCCD, tailored Coupled Cluster}

\usepackage{setspace}
\begin{document}

\begin{abstract}
\singlespacing
\linespread{1.0}
Tailored coupled cluster theory represents a computationally inexpensive way to describe static and dynamical electron correlation effects.
In this work, we scrutinize the performance of various tailored coupled cluster methods externally corrected by electronic wave functions of polynomial cost.
Specifically, we focus on frozen-pair coupled-cluster (fpCC) methods, which are externally corrected by pair-coupled cluster doubles (pCCD), and coupled cluster theory tailored by matrix product state wave functions optimized by the density matrix renormalization group (DMRG) algorithm.
As test system, we selected a set of various small- and medium-sized molecules containing diatomics (N$_2$, F$_2$, C$_2$, CN$^+$, BN, BO$^+$, and Cr$_2$) and molecules (ammonia, ethylene, cyclobutadiene, benzene) for which conventional single-reference coupled cluster singles and doubles (CCSD) is not able to produce accurate results for spectroscopic constants, potential energy surfaces, and barrier heights.
Most importantly, DMRG-tailored and pCCD-tailored approaches yield similar errors in spectroscopic constants and potential energy surfaces compared to multireference and/or experimental reference data and generally outrank the conventional single-reference CCSD approach.
Although fpCC methods provide a reliable description for the dissociation pathway of molecules featuring single and quadruple bonds, they fail in the description of triple or hextuple bond-breaking processes or avoided crossing regions. 
\end{abstract}

\singlespacing
\linespread{1.0}

\section{Introduction}

Conventional single-reference coupled cluster theory provides a robust and systematically improvable treatment of electron correlation effects.\cite{Coester1958,Cizek1966,Cizek1971,Paldus1972,Bartlett1981,MEST,MBM,Bartlett2007}
The wave function ansatz features an exponential cluster operator that ensures size-extensivity, while the straightforward truncation scheme of the cluster operator based on the excitation level yields an ordered hierarchy of approximations that converges toward the full configuration interaction limit.
The coupled cluster model in its standard single-reference formulation is one of the most accurate tools in describing dynamical electron correlation but it fails when the electronic systems under study has multireference character.
In such cases, the hierarchy of approximations breaks down and the truncation of the cluster operator provides incorrect approximations to the exact electronic wave function with unphysical coupling between cluster amplitudes.\cite{paldus1999,cc-piecuch,cc-scuseria}

One possible remedy dedicated to capture strong electron correlation effects are externally corrected or tailored coupled cluster methods\cite{externally_corrected_cc, externally_corrected_cc_2,externally_corrected_cc_3, externally_corrected_cc_4, externally_corrected_cc_5, externally_corrected_cc_6, externally_corrected_cc_8, externally_corrected_cc_9, dmrg-tcc-2016, dmrg-tcc-2019, dmrg-tcc-2020}. In this methodology, a subset of cluster amplitudes is extracted from an external model that guarantees the proper description of the multireference nature of the molecular system under study.
Popular wave function approaches to except some external coupled cluster amplitudes include the multireference configuration interaction (MRCI) or complete active space self-consistent field (CASSCF) methods.
However, these approaches are computationally infeasible for large molecules and force us to resign from black-box computational setups. 
As an alternative to conventional electronic structure methods, the density matrix renormalization group (DMRG) algorithm\cite{white, dmrg-4,  dmrg-5,  dmrg-6, dmrg-7, dmrg-8, dmrg-9} and various geminal-based approaches\cite{geminal1971, geminal1972, gvb-1973, surjan1999, apsg-2002, geminals2007, surjan2012, pccd-2013-limacher, scuseria2013, pccd-2014-prb, pccd-2014-jcp, pccd-2014-jpca, pccd-2014-jctc, pccd-2014-prc, pccd-2014-stein, seniority-cc-2014, apsg-pt-2014, bytautas2015, geminals_lcc_2015, pccd-2015, erpa-2016, geminals-2016, Nowak2019, brzek2019}
offer a computationally less complex way to model strongly-correlated electrons.

The DMRG algorithm represents a computationally efficient method to optimize matrix product state (MPS) wave functions.
The evaluation of the electronic energy scales only polynomially with system size.
Therefore, DMRG allows us to efficiently handle large active orbital spaces.
Due to its favorable computational scaling, the DMRG algorithm found numerous applications to investigate strongly-correlated systems including transition metal-\cite{marti2008, cr2_2011, fenoDMRG, kurashige2013, Corinne_2015, Zhao2015, freitag2015, freitag2015errata}
or actinide-containing molecules.~\cite{cuo_dmrg, boguslawski2017, ola-neptunyl-cci-2018}
Although being computationally more efficient, DMRG still requires us to select active space orbitals.
This can be done efficiently exploiting, for instance, tools based on quantum information theory\cite{Ziesche1995, legeza2006, rissler2006, qi-2012, dmrg-2013, vedral2014, dmrg-2015, freitag2015, freitag2015errata, dmrg-2015-ijqc,dmrg-2016, Schilling2016, autocas2016, autocas2016-2, ijqc-eratum, autocas2017, boguslawski2017,ding2020, ding2021}, fractional occupancies of unrestricted natural orbitals\cite{unocas1989, unocas1998}, or high-spin-state unrestricted Hartre--Fock (UHF) natural orbitals.~\cite{abccas2018,abccas2019}

In this work, we benchmark various coupled cluster singles and doubles models tailored by unconventional electronic structure approaches.
Specifically, we focus on orbital-optimized pair coupled cluster doubles (pCCD)~\cite{pccd-2014-prb,pccd-2014-stein} and DMRG wave functions.
The optimization of these wave functions scales only polynomially with system size.
Hence, these approaches provide an efficient way to capture strong electron correlation effects (within an active orbital space in the case of DMRG).
Most importantly, both methods allow for an accurate description of static/nondynamic electron correlation effects~\cite{qi-2012,dmrg-2016} and thus represent a promising choice for reference coupled cluster amplitudes in externally corrected coupled cluster methods.
Furthermore, the quality of DMRG calculations is affected by the type of molecular orbitals used in the active orbital space.
Recent studies report that local pair natural orbitals and domain-based local pair natural orbitals perform better than canonical RHF orbitals in DMRG-tCCSD.\cite{dmrg-tcc-2016, dmrg-tcc-2019-jcp, dmrg-tcc-2020-jctc, dmrg-tcc-2020-4c}
Here, we benchmark another type of orbitals of localized nature, namely pCCD-optimized orbitals as they have not yet been combined with coupled cluster theory tailored by DMRG wave functions.

Furthermore, the optimal active orbital space used in DMRG calculations that are then employed in the tailored coupled cluster flavour can be chosen according to the selection protocol presented by some of us~\cite{dmrg-tcc-2019}. This active space selection protocol provides accurate values for the correlation energy.
In this work, however, we focus on a different approach.
We aim at constructing a minimal but optimal active orbital space that captures the dominant part of static/nondynamic electron correlation using a one-step procedure exploiting tools based on quantum information theory.
Such an active-space selection scheme will be beneficial for large-scale modelling or the accurate and efficient prediction of potential energy surfaces as it reduces the number of DMRG calculations to be performed on the daily basis.
Besides, active orbital spaces can change along the reaction pathway. Thus, ensuring the same composition of active orbitals comprised in large active space calculation might be difficult to achieve along the reaction coordinate.

This work is organized as follows.
In Section 2, we briefly review the main ideas of tailored-coupled-cluster methods, followed by the pCCD- and DMRG-tailored flavours. 
Computational details are presented in Section 3.
We discuss the results and performance of these methods in Section 4.
Finally, we conclude in section 5.

\section{Tailored coupled cluster theory}
\label{section:theory}

The core of coupled cluster theory is the exponential parametrization of the wave function.
Tailored coupled cluster approaches take advantage of this ansatz as the exponential form allows us to utilize the Baker-Campbell-Hausdorff (BCH) formula and to factorize any operator of the form ${e}^{\hat{T}} = {e}^{\hat{T}_\mathrm{a}+\hat{T}_{b}}$ to $\mathrm{e}^{\hat{T}_\mathrm{a}}\mathrm{e}^{\hat{T}_\mathrm{b}}$ if and only if the operators $\hat{T}_\mathrm{a}$ and $\hat{T}_\mathrm{b}$ commute.
Furthermore, the particular partitioning scheme of the cluster operator (here in $\hat{T}_\mathrm{a}$ and $\hat{T}_\mathrm{b}$) depends on the external model for strong correlation.
The tailored coupled cluster wave function is thus expressed as
\begin{equation}\label{eq:tcc}
\ket{\Phi_{\rm tCC}} = e^{\hat{T}} \ket{\Phi_0} = e^{\hat{T}_\mathrm{a}} e^{\hat{T}_\mathrm{b}} \ket{\Phi_0},
\end{equation}
where $\ket{\Phi_0}$ is a reference Slater determinant and $\hat{T}$ is the cluster operator that is partitioned into a sum of two cluster operators, $\hat{T}_\mathrm{a}$ and $\hat{T}_\mathrm{b}$.
Note that we assume that $\hat{T}_\mathrm{a}$ and $\hat{T}_\mathrm{b}$ do commute.
The cluster amplitudes of one part of the composite cluster operators $\hat{T}$, here say $\hat{T}_\mathrm{b}$, are then derived from some external calculation that provides a proper treatment of strong correlation.
With the $\hat{T}_\mathrm{b}$ amplitudes being frozen, the remaining cluster amplitudes of $\hat{T}_\mathrm{a}$ can be obtained using optimization algorithms that are analogous to single-reference coupled cluster methods.
That is, the cluster amplitudes of $\hat{T}_\mathrm{a}$ can be obtained using projection techniques, where the Schr\"{o}dinger equation for this particular wave function ansatz reads
\begin{align}\label{eq:hcc}
e^{-\hat{T}_\mathrm{b}} e^{-\hat{T}_\mathrm{a}}\hat{H} e^{\hat{T}_\mathrm{a}} e^{\hat{T}_\mathrm{b}} \ket{ \Phi_0} &= E  \ket{\Phi_0}.
\end{align}
In the above equation, the $\hat{T}_\mathrm{b}$ amplitudes are kept fixed during the optimization.
Thus, the projection manifold of the tailored coupled cluster wave function contains only the set of determinants $\{\hat{T}_\mathrm{a} \ket{\Phi_0}\}$.

\subsection{Frozen-pair coupled cluster theory}\label{s:fpcc}

Frozen-pair coupled cluster theory originates from the idea of seniority-based coupled cluster approaches where the components that differ in the number of unpaired electrons are treated separately.
The most significant part of the wave function is the seniority-zero sector, which includes only those amplitudes where the number of unpaired electrons is zero \cite{pccd-2014-jcp}.
The pCCD wave function is an example for such an ansatz.
That is, the pCCD wave function is constructed from two-electron functions, also called geminals, using an exponential cluster operator of the form,
\begin{equation}
\ket{\Phi_{\rm pCCD}} = e^{\hat{T}_\mathrm{p}} \ket{\Phi_0},
\end{equation}
where the cluster operator $\hat{T}_\mathrm{p}$ contains only electron pair-excitations (geminals),
\begin{equation}\label{eq:Tp}
\hat{T}_\mathrm{p} = \sum_i^{\rm occ}\sum_{a}^{virt} t_{ii}^{aa} a_a^{\dagger}a_{\bar {a}}^{\dagger} a_{\bar{i}} a_i 
\end{equation}
with $a_p$ ($a_p^\dagger$) and $a_{\bar{p}}$ ($a_{\bar{p}}^\dagger$) being electron annihilation (creation) operators for $\alpha$- and $\beta$-spin electrons, respectively.
If combined with an orbital optimization protocol, the pCCD method is size-consistent and provides a proper qualitative description of the (exact) electronic wave function for strongly-correlated electronic systems \cite{pccd-2013-limacher, pccd-2014-prb, pccd-2014-jcp, pccd-2014-jctc, pccd-2014-prc, pccd-2014-stein}.
However, energetics and other properties cannot be described precisely (for instance, satisfying chemical accuracy) by restricting the wave function to the seniority-zero sector alone.
To reach quantitative accuracy, we need to go beyond the electron-pair approximation and to extend the electronic wave function by components that account for unpaired electrons, so-called broken-pairs~\cite{seniority-cc-2014,pccd-pt-2014,geminals_lcc_2015,garza2015actinide,pccd-PTX}.

Frozen-pair coupled cluster (fpCC) theory chooses the pCCD wave function as the fixed reference function \cite{seniority-cc-2014}.
That is, the cluster operator $\hat{T}_\mathrm{b}$ in eq.~\eqref{eq:tcc} is equivalent to the electron-pair cluster operator of eq.~\eqref{eq:Tp}
and the cluster amplitudes are thus divided to pair-amplitudes ($\hat{T}_\mathrm{b}$) and non-pair amplitudes ($\hat{T}_\mathrm{a}$).
The fpCC ansatz therefore reads
\begin{equation}
\ket{\Phi_{\rm fpCC}} = e^{\hat{T}'} \ket{\Phi_{\rm pCCD}} = e^{\hat{T}'} e^{\hat{T_\mathrm{p}}} \ket{\Phi_0},
\end{equation}
where $\hat{T}'$ is a cluster operator that is restricted to contain electron excitations (singles, broken-pair doubles, etc.) beyond electron-pair excitations.
In the fpCCD method, the cluster operator is defined as $\hat{T}' = \hat{T}_2 - \hat{T}_{\rm p}$, while for fpCCSD, the cluster operator includes also single excitations, $\hat{T}' = \hat{T}_1 + \hat{T}_2 - \hat{T}_{\rm p}$.
The geminal amplitudes $\{c_i^a\}$ account for strong electron correlation effects, while the remaining amplitudes introduce broken-pair components to complement the wave function.

The difficulties in coupled cluster theory arise from the non-linearity of the amplitude equations that have to be solved to obtain the cluster amplitudes.
This technical obstacle can be bypassed by truncating/neglecting the non-linear terms in the BCH expansion.
Although linearized coupled cluster (LCC) theory did not gain popularity due to its poor performance, the linearized version of pCCD-tailored coupled cluster approaches allows us to reach chemical accuracy for many challenging systems.~\cite{geminals_lcc_2015,pccd-PTX,filip-jctc-2019,Nowak2019,pawel-yb2}
Specifically, it has been shown that the pCCD-tailored LCC method is an efficient and reliable alternative to conventional electronic structure methods for both ground- \cite{geminals_lcc_2015,pccd-PTX,filip-jctc-2019} and excited states \cite{Nowak2019,pawel-yb2}.
The ansatz is given by
\begin{equation}
\ket{\Phi_{\rm fpLCC}} \approx (1+\hat{T}')  \ket{\Phi_{\rm pCCD}} = (1+\hat{T}') e^{\hat{T_\mathrm{p}}} \ket{\Phi_0},
\end{equation}
where $\hat{T}'$ is again some cluster operator that is restricted to contain non-pair electron excitations.
The coupled cluster equations are linear with respect to non-pair amplitudes but the coupling between all pair- and non-pair amplitudes is included (in addition to all non-linear terms originating from the pCCD reference function).
In this work, we use truncated coupled cluster models that include either only double excitations (fpLCCD) or single and double excitations (fpLCCSD).

\subsection{The DMRG-tailored coupled cluster method}

The DMRG algorithm provides qualitatively correct solutions within some active orbital space, while the wave function components that include external or inactive orbitals can be included \textit{a posteriori} using, for example, DMRG-tailored CC approaches.~\cite{dmrg-tcc-2016}
In these externally corrected CC flavours, the matrix product state ansatz, which is optimized by the DMRG algorithm, is translated to a CI-type wave function~\cite{dmrg-casci} with some specific reference determinant. Spin-dependent coupled cluster amplitudes can then be evaluated from the reconstructed CI coeffcients using standard equations,
\begin{gather}
t_i^a = c_{i}^{a} / c_0 \\
t_{ij}^{ab} = c_{ij}^{ab} / c_0 - (c_i^a c_j^b - c_i^b c_j^a)  / c_0^2,
\end{gather}
where $c_0$ is the CI coefficient of the chosen reference determinant and the indices indicate spin orbitals.
The spin-free amplitudes optimized by solving the spin-summed CC amplitude equations can be deduced form the spin-dependent ones as $t_I^A = t_{I_\alpha}^{A_\alpha}$ and $t_{IJ}^{AB} = t_{I_\alpha J_\beta}^{A_\alpha B_\beta}$, where the spin degree of freedom is labeled as a subscript and capital letters indicate spatial orbitals.
Within the DMRG-tCCSD formalism, the wave function is optimized in its split-amplitude form,
\begin{equation} \label{eq:wcc}
\ket{\Phi_{\rm DMRG-tCCSD}} = e^{\hat{T}_{\rm CAS}} e^{\hat{T}_{\rm ext}} \ket{\Phi_0},
\end{equation}
where the $\hat{T}_{\rm CAS}$ cluster operator includes amplitudes comprising excitations within the active space orbitals, while the operator $\hat{T}_{\rm ext}$ incorporates all excitations beyond the active space \cite{kinoshita2005}.
The $\hat{T}_{\rm CAS}$ amplitudes are extracted from DMRG calculations and kept frozen during the optimization of the remaining amplitudes.
That is, the $\hat{T}_{\rm ext}$ amplitudes are obtained from the solution of the conventional CCSD equations where the $\hat{T}_{\rm CAS}$ amplitudes are fixed.
Preventing the relaxation of the $\hat{T}_{\rm CAS}$ amplitudes allows us to capture strong electron correlation effects within the CCSD framework, while the relaxed $\hat{T}_{\rm ext}$ amplitudes are optimized to supplement the wave function by the missing dynamical electron correlation effects.

\section{Computational Details}
\label{section:details}

\subsection{Electronic structure calculations}

All pCCD and (tailored) coupled cluster calculations (using the spin-summed equations) were performed in a developer version of the PyBEST software package\cite{pybest-2021,pybest-1.0.0}.
We used the aug-cc-pVDZ and aug-cc-pVTZ basis sets for the F atom, the cc-pCVDZ basis set for the benzene molecule, and Dunning's cc-pVDZ and cc-pVTZ basis sets for all other atoms.
For the N$_2$ dimer, we performed additional calculations for the cc-pVQZ basis set.~\cite{basis-cc-pvdz,basis-cc-pvtz,basis-cc-pvqz,basis-cc-pcvdz}
The depths of the potential energy well were obtained from a generalized Morse function\cite{morse_potential} fit.
The vibrational frequencies and equilibrium bond lengths were calculated numerically from a polynomial fit of sixth order around the equilibrium distance, where we used $M_{\textrm{B}} = 11.0093\,u$ for the B atom, $M_{\textrm{C}} = 12\,u$ for the C atom, $M_{\textrm{N}} = 14.0031\,u$ for the N atom, $M_{\textrm{O}} = 15.9949\,u$ for the O atom, and $M_{\textrm{F}} = 18.9984\,u$ for the F atom.
The CCSD(T) and CCSDT calculations for ammonia, ethylene, and cyclobutadiene have been performed with the Molpro 2012.1.12 software package.\cite{molpro2012, molpro2012_2}

The spin-free DMRG calculations were performed using the Budapest QC-DMRG program.~\cite{dmrg_ors}
Two different sets of molecular orbitals were investigated, namely canonical restricted Hartree-Fock (RHF) orbitals and pCCD-optimized orbitals.
Furthermore, we aimed at constructing small but chemically reasonable active orbital spaces as they allow us to maintain the same level of approximation for orbitals of similar weight and correlation strength.
Specifically, it allows us to avoid errors driven by an unbalanced composition of the active space.
The active space selection for RHF orbitals was based on quantum information measures obtained from $m = {64, 128}$ DMRG calculations since DMRG calculations exploiting even small bond dimensions provide robust single-orbital entropies and orbital-pair mutual information profiles.

In all calculations involving active orbital spaces, we performed calculations with different values of the bond dimension $m$ to ensure that DMRG energy converged with respect to $m$, that is, we used  $m={32, 64, 128}$ for all first-row atom dimers, $m={128, 256, 512}$ for Cr$_2$, ammonia, ethylene, and the benzene molecules, and $m={256, 512, 1024}$ for cyclobutadiene molecule.
The DMRG energies of all first-row atom dimers, ammonia, and benzene were converged up to $\Delta E = 10^{-8}$ with respect to $m$, while for the Cr$_2$, ethylene and cyclobutadiene molecules the convergence threshold was relaxed to $\Delta E = 10^{-5}$ and $\Delta E = 10^{-4}$, respectively.
The converged DMRG wave functions were first used to reconstruct the CI and then CC coefficients.

The active space of the N$_2$ and F$_2$ molecules consists of one $3\sigma_g$, two $\pi_u$, two $\pi_g$, and one $3 \sigma_u$ orbital.
In the case of the carbon dimer, the active orbital space was extended by the $2\sigma_g$ and $2\sigma_u$ orbitals as the single orbital entropy and orbital-pair mutual information in the pCCD-optimized orbital basis suggested that these orbitals might have non-negligible impact on the balanced description of electron correlation effects in the dissociation region.
The orbital interactions are similar for the C$_2$ isoelectronic analogues --- BN, BO$^+$, and CN$^+$ --- and, therefore, their active orbital spaces were composed of two $\sigma$, two $\sigma^*$, two $\pi$, and two $\pi^*$ orbitals occupied by eight electrons.
For the chromium dimer, we used all twelve valence orbitals (4s and 3d) following the recommendations of Refs.~\citenum{cr2_2016, dmrg-tcc-2016}.

We studied two active spaces in the case of the ammonia compound, which were selected solely on the values of the single orbital entropies and the orbital-pair mutual information.
Specifically, we looked for pronounced gaps in the single-orbital entropy and orbital-pair mutual information distributions to find reasonable cutoff values.
This procedure resulted in a small CAS(6,6) containing orbitals with $s_i$ > 0.04 and $I_{ij}$ > 0.07, while a slightly bigger CAS(8,8) was derived from a decreased cutoff value of $s_i$ > 0.03 and $I_{ij}$ > 0.01.
A similar active space selection protocol was used to obtain a CAS(12,12) for the ethylene molecule and a CAS(20,20) for the cyclobutadiene complex.

For benzene, the thresholds was tightened up ($I_{ij}$ > 0.1) resulting in a CAS(6,6).
The diagrams for the single-orbital entropy and orbital-pair mutual information of selected systems (molecules and bond lengths) are summarized in the Supporting Information (SI).
The coupled cluster amplitudes were reconstructed from matrix product state wave functions using the method described by some of us.\cite{dmrg-tcc-2016}
All tailored coupled cluster calculations were performed using the CC amplitude equations in the spin-summed form.

\subsection{Abbreviations of method names}

In all (conventional and tailored) coupled cluster calculations, we used two different reference wave functions and molecular orbitals: (a) canonical RHF and (b) (variationally) orbital-optimized pCCD. 
In this work, all coupled cluster methods were truncated at the doubles (CCD) and singles and doubles (CCSD) levels.
Thus, CCD$^a$ and CCSD$^a$ represent the traditional coupled cluster methods with a canonical RHF reference function, while the abbreviations CCD$^b$ and CCSD$^b$ indicate that the reference determinant (and hence the molecular orbitals) of the orbital-optimized pCCD wave function was selected as the reference determinant in conventional, that is untailored, coupled cluster calculations.
In these flavours, all cluster amplitudes are thus optimized and all information about the electron-pair amplitudes is lost.
We should note that the linearized CC corrections with a pCCD reference function are labeled as fpLCCD and fpLCCSD, respectively, while they were originally introduced using the acronyms pCCD-LCCD and pCCD-LCCSD.
All fpCC calculations, including the linearized variants, have been performed in the pCCD-optimized orbital basis only.

\section{Results}
\label{section:results}

\begin{table}[H]
\caption{Spectroscopic constants for the dissociation of homonuclear main-group diatomic molecules for different quantum chemistry methods and basis sets.
Errors with respect to MRCI, FCIQMC, or DMRG are given in parentheses.
The reference data has no error given in parentheses.
The superscript $a$ denotes that calculations have been performed in the canonical RHF orbital basis, while the superscript $b$ stands for pCCD-optimized orbitals.
The ``$*$'' denotes the lowest-lying singlet excited state.
}
\label{tab:3-1}
\begin{tiny}
\begin{tabular}{p{0.01\textwidth}p{0.15\textwidth} p{0.02\textwidth}r p{0.02\textwidth}r p{0.02\textwidth}r p{0.02\textwidth}r p{0.02\textwidth}r p{0.02\textwidth}r}
	&		&	\multicolumn{2}{l}{	r$_\mathrm{e}$ [\r{A}]				}	&	\multicolumn{2}{l}{	D$_\mathrm{e}$ [$\rm \frac{kcal}{mol}$]				}	&	\multicolumn{2}{l}{	$\omega_\mathrm{e}$ [cm$^{-1}$]				}	&	\multicolumn{2}{l}{	r$_\mathrm{e}$ [\r{A}]				}	&	\multicolumn{2}{l}{	D$_\mathrm{e}$ [$\rm \frac{kcal}{mol}$]				}	&	\multicolumn{2}{l}{	$\omega_\mathrm{e}$ [cm$^{-1}$]				}	\\	\hline
	&		&	\multicolumn{6}{l}{aug-cc-pVDZ}																				&	\multicolumn{4}{l}{aug-cc-pVTZ}																				\\	\hline
F$_2$	&	RHF	&	1.338		&	   (	0.115	)	&	181.0		&	   (	-152.5	)	&	1216		&	   (	-415	)	&	1.328		&	   (	0.092	)	&			&				&	1271		&	   (	-379	)	\\	
	&	CCD$^a$	&	1.414		&	   (	0.039	)	&	69.1		&	   (	-40.6	)	&	964		&	   (	-163	)	&	1.383		&	   (	0.037	)	&	82.2		&	   (	-48.3	)	&	1053		&	   (	-161	)	\\	
	&	CCSD$^a$	&	1.425		&	   (	0.028	)	&	57.0		&	   (	-28.5	)	&	922		&	   (	-121	)	&	1.392		&	   (	0.028	)	&	70.1		&	   (	-36.2	)	&	1018		&	   (	-126	)	\\	
	&	pCCD	&	1.502		&	   (	-0.049	)	&	12.6		&	   (	15.9	)	&	622		&	   (	179	)	&	1.466		&	   (	-0.046	)	&	16.2		&	   (	17.7	)	&	708		&	   (	184	)	\\	
	&	CCD$^b$	&	1.425		&	   (	0.028	)	&	55.5		&	   (	-27.0	)	&	924		&	   (	-123	)	&	1.391		&	   (	0.029	)	&	68.8		&	   (	-34.9	)	&	1021		&	   (	-129	)	\\	
	&	CCSD$^b$	&	1.424		&	   (	0.029	)	&	55.9		&	   (	-27.4	)	&	926		&	   (	-125	)	&	1.391		&	   (	0.029	)	&	69.3		&	   (	-35.4	)	&	1022		&	   (	-130	)	\\	
	&	fpLCCD	&	1.473		&	   (	-0.020	)	&	37.5		&	   (	-9.0	)	&	776		&	   (	25	)	&	1.434		&	   (	-0.014	)	&	44.3		&	   (	-10.4	)	&	866		&	   (	26	)	\\	
	&	fpLCCSD	&	1.469		&	   (	-0.016	)	&	38.1		&	   (	-9.6	)	&	788		&	   (	13	)	&	1.431		&	   (	-0.011	)	&	45.2		&	   (	-11.3	)	&	879		&	   (	13	)	\\	
	&	fpCCD	&	1.473		&	   (	-0.020	)	&	36.7		&	   (	-8.2	)	&	773		&	   (	28	)	&	1.433		&	   (	-0.013	)	&	43.3		&	   (	-9.4	)	&	863		&	   (	29	)	\\	
	&	fpCCSD	&	1.469		&	   (	-0.016	)	&	37.1		&	   (	-8.6	)	&	782		&	   (	19	)	&	1.431		&	   (	-0.011	)	&	44.2		&	   (	-10.3	)	&	873		&	   (	19	)	\\	
	&	DMRG(8,8)-tCCSD$^a$	&	1.493		&	   (	-0.040	)	&	37.7		&	   (	-9.2	)	&	728		&	   (	73	)	&	1.454		&	   (	-0.034	)	&	43.7		&	   (	-9.8	)	&	758		&	   (	134	)	\\	
	&	DMRG(8,8)-tCCSD$^b$	&	1.468		&	   (	-0.015	)	&	38.4		&	   (	-9.9	)	&	728		&	   (	73	)	&	1.427		&	   (	-0.007	)	&	45.9		&	   (	-12.0	)	&	875		&	   (	17	)	\\	
	&	MRCI\cite{peterson1993} 	&	1.453		&				&	28.5		&				&	801		&				&	1.420		&				&	33.9		&				&	892		&				\\	
	&	exp.\cite{MolSpectraStruct, irikura2007} 	&	1.412		&	   (	0.041	)	&	37.7		&	   (	-9.2	)	&	917		&	   (	-116	)	&	1.412		&	   (	0.008	)	&	37.7		&	   (	-3.8	)	&	917		&	   (	-25	)	\\	\cline{2-14}
	&		&	\multicolumn{6}{l}{cc-pVDZ}																				&	\multicolumn{4}{l}{cc-pVTZ}																				\\	\cline{2-14}
	&	pCCD	&	1.471		&	   (	-0.006	)	&	15.5		&	   (	10.8	)	&	693		&	   (	64	)	&	1.471		&	   (	-0.052	)	&	15.5		&	   (	17.0	)	&	693		&	   (	198	)	\\	
	&	CCD$^b$	&	1.392		&	   (	0.074	)	&	66.1		&	   (	-39.8	)	&	1017		&	   (	-261	)	&	1.392		&	   (	0.028	)	&	66.2		&	   (	-33.7	)	&	1016		&	   (	-126	)	\\	
	&	CCSD$^b$	&	1.391		&	   (	0.074	)	&	66.5		&	   (	-40.2	)	&	1019		&	   (	-263	)	&	1.391		&	   (	0.028	)	&	66.6		&	   (	-34.1	)	&	1018		&	   (	-128	)	\\	
	&	fpLCCD	&	1.432		&	   (	0.034	)	&	42.0		&	   (	-15.7	)	&	868		&	   (	-112	)	&	1.432		&	   (	-0.013	)	&	42.1		&	   (	-9.6	)	&	867		&	   (	23	)	\\	
	&	fpLCCSD	&	1.429		&	   (	0.036	)	&	43.1		&	   (	-16.8	)	&	879		&	   (	-123	)	&	1.429		&	   (	-0.010	)	&	43.2		&	   (	-10.7	)	&	878		&	   (	13	)	\\	
	&	fpCCD	&	1.432		&	   (	0.034	)	&	41.0		&	   (	-14.7	)	&	864		&	   (	-108	)	&	1.432		&	   (	-0.012	)	&	41.2		&	   (	-8.7	)	&	864		&	   (	27	)	\\	
	&	fpCCSD	&	1.429		&	   (	0.036	)	&	41.9		&	   (	-15.6	)	&	874		&	   (	-117	)	&	1.429		&	   (	-0.010	)	&	42.1		&	   (	-9.6	)	&	872		&	   (	18	)	\\	
	&	DMRG(8,8)-tCCSD$^a$	&	1.471		&	   (	-0.006	)	&	35.0		&	   (	-8.7	)	&	765		&	   (	-9	)	&	1.426		&	   (	-0.007	)	&	43.8		&	   (	-11.3	)	&	878		&	   (	12	)	\\	
	&	DMRG(8,8)-tCCSD$^b$	&	1.474		&	   (	-0.006	)	&	35.8		&	   (	-9.5	)	&	742		&	   (	15	)	&	1.426		&	   (	-0.007	)	&	43.4		&	   (	-10.9	)	&	872		&	   (	19	)	\\	
&	MRCI\cite{peterson1993} 	&	1.465	&				&	26.3	&				&	756	&				&	1.419	&				&	32.5	&				&	891	&				\\	
&	exp.\cite{MolSpectraStruct, irikura2007} 	&	1.412	&	   (	0.053	)	&		&				&	917	&	   (	-161	)	&	1.412	&	   (	0.007	)	&		&				&	917	&	   (	-27	)	\\	\hline
	&		&	\multicolumn{6}{l}{cc-pVDZ}																				&	\multicolumn{4}{l}{cc-pVTZ}																				\\	\hline
N$_2$	&	pCCD	&	1.099	&	   (	0.020	)	&	239.6	&	   (	-37.6	)	&	2482	&	   (	-153	)	&	1.085	&	   (	0.019	)	&	244.8	&	   (	-26.9	)	&	2517	&	   (	-176	)	\\	
	&	CCD$^b$	&	1.111	&	   (	0.008	)	&		&				&	2415	&	   (	-86	)	&	1.093	&	   (	0.011	)	&		&				&	2447	&	   (	-106	)	\\	
	&	CCSD$^b$	&	1.112	&	   (	0.008	)	&		&				&	2412	&	   (	-84	)	&	1.093	&	   (	0.011	)	&		&				&	2444	&	   (	-104	)	\\	
	&	fpLCCD	&	1.119	&	   (	0.001	)	&		&				&	2320	&	   (	8	)	&	1.101	&	   (	0.003	)	&		&				&	2347	&	   (	-7	)	\\	
	&	fpLCCSD	&	1.120	&	   (	0.000	)	&		&				&	2315	&	   (	14	)	&	1.102	&	   (	0.002	)	&		&				&	2340	&	   (	0	)	\\	
	&	fpCCD	&	1.118	&	   (	0.002	)	&		&				&	2335	&	   (	-7	)	&	1.100	&	   (	0.004	)	&		&				&	2365	&	   (	-24	)	\\	
	&	fpCCSD	&	1.118	&	   (	0.002	)	&		&				&	2336	&	   (	-8	)	&	1.100	&	   (	0.004	)	&		&				&	2365	&	   (	-24	)	\\	
	&	DMRG(6,6)-tCCSD$^a$	&	1.118	&	   (	0.002	)	&	208.1	&	   (	-6.1	)	&	2327	&	   (	2	)	&	1.100	&	   (	0.004	)	&	227.4	&	   (	-9.5	)	&	2356	&	   (	-15	)	\\	
	&	DMRG(6,6)-tCCSD$^b$	&	1.122	&	   (	-0.002	)	&	219.0	&	   (	-17.0	)	&	2383	&	   (	-55	)	&	1.102	&	   (	0.002	)	&	239.3	&	   (	-21.4	)	&	2387	&	   (	-46	)	\\	
	&	MRCI \cite{peterson1993}	&	1.120	&				&	202.0	&				&	2329	&				&	1.104	&				&	217.9	&				&	2341	&				\\	\cline{2-14}
	&		&	\multicolumn{6}{l}{cc-pVQZ}																	&	\multicolumn{4}{l}{aug-cc-pVTZ}																	\\	\cline{2-14}
	&	pCCD	&	1.077	&	   (	0.023	)	&	252.8	&	   (	-28.6	)	&	2723	&	   (	-371	)	&	1.087	&	   (	0.011	)	&	256.1	&	   (	-31.0	)	&	2450	&	   (	-91	)	\\	
	&	CCD$^b$	&	1.090	&	   (	0.010	)	&		&				&	2447	&	   (	-95	)	&	1.092	&	   (	0.006	)	&		&				&	2455	&	   (	-96	)	\\	
	&	CCSD$^b$	&	1.090	&	   (	0.010	)	&		&				&	2450	&	   (	-98	)	&	1.092	&	   (	0.006	)	&		&				&	2448	&	   (	-89	)	\\	
	&	fpLCCD	&	1.098	&	   (	0.003	)	&		&				&	2370	&	   (	-18	)	&	1.100	&	   (	-0.002	)	&		&				&	2357	&	   (	2	)	\\	
	&	fpLCCSD	&	1.098	&	   (	0.002	)	&		&				&	2373	&	   (	-21	)	&	1.101	&	   (	-0.003	)	&		&				&	2344	&	   (	15	)	\\	
	&	fpCCD	&	1.096	&	   (	0.005	)	&		&				&	2394	&	   (	-43	)	&	1.099	&	   (	-0.001	)	&		&				&	2374	&	   (	-15	)	\\	
	&	fpCCSD	&	1.096	&	   (	0.005	)	&		&				&	2402	&	   (	-51	)	&	1.099	&	   (	-0.001	)	&		&				&	2369	&	   (	-10	)	\\	
	&	DMRG(6,6)-tCCSD$^a$	&	1.097	&	   (	0.004	)	&	230.9	&	   (	-6.7	)	&	2353	&	   (	-2	)	&	1.094	&	   (	0.004	)	&	222.1	&	   (	3.0	)	&	2280	&	   (	79	)	\\	
	&	DMRG(6,6)-tCCSD$^b$	&	1.099	&	   (	0.001	)	&	244.4	&	   (	-20.2	)	&	2392	&	   (	-40	)	&	1.117	&	   (	-0.019	)	&	237.0	&	   (	-11.9	)	&	2232	&	   (	127	)	\\	
	&	MRCI\cite{peterson1993} 	&	1.101	&				&	224.2	&				&	2352	&				&		&				&		&				&		&				\\	
	&	exp.\cite{MolSpectraStruct, shimanouchi1997} 	&	1.098	&	   (	0.002	)	&	225.1	&	   (	-0.9	)	&	2359	&	   (	-7	)	&	1.098	&				&	225.1	&				&	2359	&				\\	\hline
	&		&	\multicolumn{6}{l}{cc-pVDZ}																				&	\multicolumn{4}{l}{cc-pVTZ}																				\\	\hline
C$_2$	&	CCD$^b$	&	1.263		&	   (	0.009	)	&	136.0		&	   (	-9.2	)	&	1873		&	   (	-62	)	&	1.240		&	   (	0.013	)	&	156.1		&	   (	-20.8	)	&	1911		&	   (	-76	)	\\	
	&	CCSD$^b$	&	1.263		&	   (	0.009	)	&	136.5		&	   (	-9.7	)	&	1873		&	   (	-62	)	&	1.240		&	   (	0.013	)	&	156.6		&	   (	-21.2	)	&	1915		&	   (	-80	)	\\	
	&	fpLCCD	&	1.264		&	   (	0.008	)	&	131.3		&	   (	-4.5	)	&	1901		&	   (	-90	)	&	1.241		&	   (	0.011	)	&	136.7		&	   (	-1.3	)	&	1871		&	   (	-36	)	\\	
	&	fpLCCSD	&	1.265		&	   (	0.008	)	&	134.2		&	   (	-7.4	)	&	1890		&	   (	-79	)	&	1.241		&	   (	0.011	)	&	140.0		&	   (	-4.6	)	&	1862		&	   (	-27	)	\\	
	&	fpCCD	&	1.261		&	   (	0.012	)	&	130.4		&	   (	-3.6	)	&	1905		&	   (	-94	)	&	1.237		&	   (	0.015	)	&	143.4		&	   (	-8.0	)	&	1901		&	   (	-67	)	\\	
	&	fpCCSD	&	1.260		&	   (	0.013	)	&	131.7		&	   (	-4.9	)	&	1906		&	   (	-95	)	&	1.236		&	   (	0.017	)	&	145.1		&	   (	-9.7	)	&	1905		&	   (	-70	)	\\	
	&	DMRG(8,8)-tCCSD$^a$	&	1.266		&	   (	0.007	)	&	134.3		&	   (	-7.5	)	&	1850		&	   (	-39	)	&	1.242		&	   (	0.011	)	&	147.5		&	   (	-12.1	)	&	1893		&	   (	-58	)	\\	
	&	DMRG(8,8)-tCCSD$^b$	&	1.265		&	   (	0.008	)	&	136.0		&	   (	-9.2	)	&	1899		&	   (	-88	)	&	1.249		&	   (	0.004	)	&	146.3		&	   (	-11.0	)	&	1885		&	   (	-51	)	\\	
	&	MRCI\cite{peterson1995}	&	1.251		&	   (	0.021	)	&	135.4		&	   (	-8.6	)	&	1873		&	   (	-62	)	&	1.252		&	   (	0.001	)	&	140.4		&	   (	-5.0	)	&	1840		&	   (	-5	)	\\	
	&	DMRG(12,28)\cite{wouters2014} 	&	1.272		&	   (	0.001	)	&	130.1		&	   (	-3.3	)	&	1816		&	   (	-5	)	&			&				&			&				&			&				\\	
	&	FCIQMC\cite{booth2011} 	&	1.273		&				&	126.8		&				&	1811		&				&	1.253		&				&	135.4		&				&	1835		&	  	\\	
	&	exp.\cite{MolSpectraStruct}	&	1.243		&	   (	0.030	)	&	147.8		&	   (	-21.0	)	&	1855		&	   (	-44	)	&	1.243		&	   (	0.010	)	&	147.8		&	   (	-12.4	)	&	1855		&	   (	-20	)	\\	\hline
	&		&	\multicolumn{6}{l}{cc-pVDZ}																				&	\multicolumn{4}{l}{cc-pVTZ}																				\\	\hline
C$_2^*$	&	CCD$^b$	&	1.415		&	   (	-0.005	)	&	111.3		&	   (	-23.8	)	&	1641		&	   (	-274	)	&	1.393		&	   (	-0.016	)	&	133.4		&	   (	-32.2	)	&	1970		&	   (	-546	)	\\	
	&	CCSD$^b$	&	1.416		&	   (	-0.006	)	&	111.3		&	   (	-23.8	)	&	1648		&	   (	-281	)	&	1.320		&	   (	0.057	)	&	131.9		&	   (	-30.7	)	&	1387		&	   (	37	)	\\	
	&	fpLCCD	&	1.466		&	   (	-0.056	)	&	87.5		&	   (	0.0	)	&	1421		&	   (	-54	)	&	1.404		&	   (	-0.026	)	&	94.9		&	   (	6.3	)	&	1389		&	   (	35	)	\\	
	&	fpLCCSD	&	1.449		&	   (	-0.039	)	&	91.1		&	   (	-3.6	)	&	1317		&	   (	50	)	&	1.406		&	   (	-0.028	)	&	98.4		&	   (	2.8	)	&	1348		&	   (	76	)	\\	
	&	fpCCD	&	1.425		&	   (	-0.015	)	&	91.9		&	   (	-4.4	)	&	1286		&	   (	80	)	&	1.394		&	   (	-0.016	)	&	106.7		&	   (	-5.5	)	&	1379		&	   (	45	)	\\	
	&	fpCCSD	&	1.426		&	   (	-0.016	)	&	94.4		&	   (	-6.9	)	&	1301		&	   (	66	)	&	1.395		&	   (	-0.018	)	&	109.4		&	   (	-8.2	)	&	1370		&	   (	54	)	\\	
	&	DMRG(12,28)\cite{wouters2014} 	&	1.410		&				&	87.6		&				&	1367		&				&			&				&			&				&			&				\\	
	&	exp.\cite{douay1988} 	&	1.377		&	   (	0.033	)	&	101.2		&	   (	-13.6	)	&	1424		&	   (	-57	)	&	1.377		&		&	101.2		&				&	1424		&				\\	\hline
\end{tabular}
\end{tiny}
\end{table}

\begin{table}[tpb]
\caption{Spectroscopic constants for the dissociation of main-group heteronuclear diatomic molecules for different quantum chemistry
methods and basis sets.
Errors with respect to MRCI, CMRCI+Q, FCIQMC, or experiment are given in parentheses.
The superscript $a$ denotes that calculations have been performed in the canonical RHF orbital basis, while the superscript $b$ stands for pCCD-optimized orbitals.
The ``$*$'' denotes the lowest-lying singlet excited state that could be optimized within pCCD.
}
\label{tab:3-2}
\begin{tiny}
\begin{tabular}{p{0.025\textwidth}p{0.15\textwidth} p{0.03\textwidth}r p{0.03\textwidth}r p{0.03\textwidth}r p{0.03\textwidth}r p{0.03\textwidth}r p{0.03\textwidth}r} \hline
	&		&	\multicolumn{2}{l}{	r$_\mathrm{e}$ [\r{A}]				}	&	\multicolumn{2}{l}{	D$_\mathrm{e}$ [$\rm \frac{kcal}{mol}$]				}	&	\multicolumn{2}{l}{	$\omega_\mathrm{e}$ [cm$^{-1}$]				}	&	\multicolumn{2}{l}{	r$_\mathrm{e}$ [\r{A}]				}	&	\multicolumn{2}{l}{	D$_\mathrm{e}$ [$\rm \frac{kcal}{mol}$]				}	&	\multicolumn{2}{l}{	$\omega_\mathrm{e}$ [cm$^{-1}$]				}	\\	\hline
	&		&	\multicolumn{6}{l}{cc-pVDZ}		&	\multicolumn{4}{l}{cc-pVTZ}									\\	\hline
BO$^+$	&	CCD$^a$	&	1.199		&	   (	0.027	)	&	114.8		&	   (	3.2	)	&	1901		&	   (	-168	)	&	1.186		&	   (	0.030	)	&	126.5		&	   (	-0.5	)	&	1966		&	   (	-182	)	\\	
	&	CCSD$^a$	&	1.205		&	   (	0.021	)	&	119.6		&	   (	-1.6	)	&	1883		&	   (	-150	)	&	1.191		&	   (	0.025	)	&	130.6		&	   (	-4.6	)	&	1949		&	   (	-165	)	\\	
	&	CCD$^b$	&	1.198		&	   (	0.028	)	&	108.5		&	   (	9.6	)	&	1903		&	   (	-169	)	&	1.184		&	   (	0.031	)	&	122.8		&	   (	3.2	)	&	2005		&	   (	-221	)	\\	
	&	CCSD$^b$	&	1.205		&	   (	0.021	)	&	111.5		&	   (	6.6	)	&	1894		&	   (	-160	)	&	1.191		&	   (	0.025	)	&	124.3		&	   (	1.7	)	&	1952		&	   (	-168	)	\\	
	&	fpLCCD	&	1.205		&	   (	0.021	)	&	107.5		&	   (	10.6	)	&	1854		&	   (	-121	)	&	1.191		&	   (	0.025	)	&	116.3		&	   (	9.7	)	&	1942		&	   (	-158	)	\\	
	&	fpLCCSD	&	1.232		&	   (	-0.006	)	&	120.5		&	   (	-2.5	)	&	2264		&	   (	-530	)	&	1.245		&	   (	-0.029	)	&	132.0		&	   (	-6.0	)	&	1865		&	   (	-81	)	\\	
	&	fpCCD	&	1.203		&	   (	0.023	)	&	106.1		&	   (	12.0	)	&	1874		&	   (	-140	)	&	1.189		&	   (	0.026	)	&	114.8		&	   (	11.3	)	&	1958		&	   (	-174	)	\\	
	&	fpCCSD	&	1.205		&	   (	0.021	)	&	107.0		&	   (	11.0	)	&	1925		&	   (	-191	)	&	1.193		&	   (	0.023	)	&	114.4		&	   (	11.6	)	&	1936		&	   (	-152	)	\\	
	&	DMRG(8,8)-tCCSD$^a$	&	1.230		&	   (	-0.004	)	&	130.8		&	   (	-12.8	)	&	1823		&	   (	-89	)	&	1.218		&	   (	-0.002	)	&	141.7		&	   (	-15.7	)	&	1868		&	   (	-84	)	\\	
	&	DMRG(8,8)-tCCSD$^b$	&	1.188		&	   (	0.038	)	&	102.0		&	   (	16.0	)	&	1676		&	   (	58	)	&	1.166		&	   (	0.049	)	&	93.3		&	   (	32.7	)	&	1718		&	   (	66	)	\\	
	&	CASSCF\cite{peterson1995} 	&	1.211		&	   (	0.016	)	&	133.7		&	   (	-15.7	)	&	1815		&	   (	-82	)	&	1.205		&	   (	0.011	)	&	138.5		&	   (	-12.4	)	&	1835		&	   (	-52	)	\\	
	&	CMRCI\cite{peterson1995} 	&	1.225		&	   (	0.001	)	&	118.6		&	   (	-0.5	)	&	1741		&	   (	-7	)	&	1.214		&	   (	0.001	)	&	126.7		&	   (	-0.7	)	&	1792		&	   (	-8	)	\\	
	&	CMRCI+Q\cite{peterson1995} 	&	1.226		&				&	118.0		&				&	1734		&				&	1.216		&				&	126.0		&				&	1784		&				\\	\hline
	&		&	\multicolumn{6}{l}{cc-pVDZ}																				&	\multicolumn{4}{l}{cc-pVTZ}																				\\	\hline
BN	&	pCCD	&	1.269		&	   (	0.030	)	&	100.3		&	   (	48.2	)	&	1776		&	   (	-125	)	&			&	   (	1.285	)	&			&	   (	154.4	)	&			&	   (	1682	)	\\	
	&	CCD$^b$	&	1.293		&	   (	0.005	)	&	150.8		&	   (	-2.2	)	&	1689		&	   (	-39	)	&	1.279		&	   (	0.006	)	&	164.4		&	   (	-10.1	)	&	1683		&	   (	-1	)	\\	
	&	CCSD$^b$	&	1.288		&	   (	0.010	)	&	153.3		&	   (	-4.7	)	&	1714		&	   (	-63	)	&	1.271		&	   (	0.014	)	&	167.7		&	   (	-13.3	)	&	1759		&	   (	-77	)	\\	
	&	fpLCCD	&	1.294		&	   (	0.004	)	&	148.5		&	   (	0.1	)	&	1683		&	   (	-33	)	&	1.281		&	   (	0.004	)	&	187.7		&	   (	-33.3	)	&	1461		&	   (	221	)	\\	
	&	fpLCCSD	&	1.293		&	   (	0.005	)	&	154.3		&	   (	-5.8	)	&	1695		&	   (	-45	)	&	1.297		&	   (	-0.012	)	&	193.9		&	   (	-39.5	)	&	1595		&	   (	87	)	\\	
	&	fpCCD	&	1.291		&	   (	0.008	)	&	146.2		&	   (	2.4	)	&	1700		&	   (	-49	)	&	1.282		&	   (	0.003	)	&	184.0		&	   (	-29.7	)	&	1589		&	   (	93	)	\\	
	&	fpCCSD	&	1.287		&	   (	0.012	)	&	147.8		&	   (	0.8	)	&	1724		&	   (	-74	)	&	1.276		&	   (	0.009	)	&	185.7		&	   (	-31.4	)	&	1606		&	   (	76	)	\\	
	&	DMRG(8,8)-tCCSD$^a$	&	1.293		&	   (	0.005	)	&	159.6		&	   (	-11.0	)	&	1787		&	   (	-136	)	&	1.282		&	   (	0.003	)	&	169.7		&	   (	-15.4	)	&	1836		&	   (	-154	)	\\	
	&	DMRG(8,8)-tCCSD$^b$	&	1.297		&	   (	0.001	)	&	138.7		&	   (	9.9	)	&	1742		&	   (	-91	)	&	1.298		&	   (	-0.013	)	&	159.2		&	   (	-4.8	)	&	1852		&	   (	-170	)	\\	
	&	CASSCF\cite{peterson1995} 	&	1.294		&	   (	0.004	)	&	158.8		&	   (	-10.2	)	&	1681		&	   (	-30	)	&	1.288		&	   (	-0.003	)	&	160.8		&	   (	-6.4	)	&	1686		&	   (	-4	)	\\	
	&	CMRCI\cite{peterson1995} 	&	1.298		&	   (	0.001	)	&	149.8		&	   (	-1.2	)	&	1655		&	   (	-4	)	&	1.284		&	   (	0.001	)	&	155.9		&	   (	-1.5	)	&	1687		&	   (	-5	)	\\	
	&	CMRCI+Q\cite{peterson1995} 	&	1.298		&				&	148.6		&				&	1651		&				&	1.285		&				&	154.4		&				&	1682		&				\\	\hline
	&		&	\multicolumn{6}{l}{cc-pVDZ}																				&	\multicolumn{4}{l}{cc-pVTZ}																				\\	\hline
    CN$^+$	&	CCD$^b$	&	1.186		&	   (	0.012	)	&	173.6		&	   (	-9.8	)	&			&				&	1.186		&	   (	0.012	)	&	173.6		&	   (	-9.8	)	&	2106		&	   (	-127	)	\\	
	&	CCSD$^b$	&	1.184		&	   (	0.014	)	&	173.0		&	   (	-9.2	)	&			&				&	1.184		&	   (	0.014	)	&	173.0		&	   (	-9.2	)	&	2121		&	   (	-141	)	\\	
	&	fpLCCD	&	1.174		&	   (	0.024	)	&	179.5		&	   (	-15.7	)	&	2068		&	   (	-88	)	&	1.174		&	   (	0.024	)	&	179.5		&	   (	-15.7	)	&	2068		&	   (	-88	)	\\	
	&	fpLCCSD	&	1.173		&	   (	0.025	)	&	176.3		&	   (	-12.5	)	&	2080		&	   (	-100	)	&	1.173		&	   (	0.025	)	&	176.3		&	   (	-12.5	)	&	2080		&	   (	-100	)	\\	
	&	fpCCD	&	1.169		&	   (	0.029	)	&	167.5		&	   (	-3.7	)	&	2075		&	   (	-95	)	&	1.169		&	   (	0.029	)	&	167.5		&	   (	-3.7	)	&	2070		&	   (	-91	)	\\	
	&	fpCCSD	&	1.181		&	   (	0.017	)	&	176.8		&	   (	-13.0	)	&	2081		&	   (	-101	)	&	1.160		&	   (	0.038	)	&	176.8		&	   (	-13.0	)	&	2174		&	   (	-194	)	\\	
	&	DMRG(8,8)-tCCSD$^a$	&	1.191		&	   (	0.007	)	&	171.2		&	   (	-7.4	)	&	2013		&	   (	-34	)	&	1.179		&	   (	0.019	)	&	155.3		&	   (	8.5	)	&	2156		&	   (	-177	)	\\	
	&	DMRG(8,8)-tCCSD$^b$	&	1.190		&	   (	0.008	)	&	178.5		&	   (	-14.7	)	&	1986		&	   (	-7	)	&	1.170		&	   (	0.028	)	&	170.9		&	   (	-7.0	)	&	2042		&	   (	-62	)	\\	
	&	CASSCF\cite{peterson1995} 	&	1.191		&	   (	0.007	)	&	178.8		&	   (	-14.9	)	&	2030		&	   (	-51	)	&	1.182		&	   (	0.016	)	&	181.0		&	   (	-17.2	)	&	2018		&	   (	-39	)	\\	
	&	CMRCI\cite{peterson1995} 	&	1.197		&	   (	0.001	)	&	165.2		&	   (	-1.4	)	&	1985		&	   (	-6	)	&	1.182		&	   (	0.016	)	&	170.6		&	   (	-6.8	)	&	2006		&	   (	-26	)	\\	
	&	CMRCI+Q\cite{peterson1995} 	&	1.198		&				&	163.8		&				&	1979		&				&	1.198		&				&	163.8		&				&	1979		&				\\	
	&	exp.\cite{MolSpectraStruct}	&	1.173		&	   (	0.025	)	&			&				&	2033		&	   (	-54	)	&	1.173		&	   (	0.025	)	&			&				&	2033		&	   (	-54	)	\\	\hline
	&		&	\multicolumn{6}{l}{cc-pVDZ}																				&	\multicolumn{4}{l}{cc-pVTZ}																				\\	\hline
(CN$^+$)$^*$	&	CCD$^b$	&	1.335		&				&	160.7		&				&	1962		&				&	1.335		&				&	160.7		&				&	1962		&				\\	
	&	CCSD$^b$	&	1.339		&				&	162.8		&				&	2084		&				&	1.339		&				&	162.8		&				&	2084		&				\\	
	&	fpLCCD	&	1.389		&				&	123.0		&				&	1099		&				&	1.389		&				&	123.0		&				&	1099		&				\\	
	&	fpLCCSD	&	1.391		&				&	142.8		&				&	1407		&				&	1.391		&				&	142.8		&				&	1407		&				\\	
	&	fpCCD	&	1.375		&				&	128.0		&				&	1264		&				&	1.375		&				&	128.0		&				&	1363		&				\\	
	&	fpCCSD	&	1.373		&				&	144.7		&				&	1515		&				&	1.373		&				&	144.7		&				&	1510		&				\\	\hline
	&		&	\multicolumn{6}{l}{cc-pVDZ}																				&	\multicolumn{4}{l}{cc-pVTZ}																				\\	\hline
CO	&	CCD$^a$	&	1.135	&	   (	0.010	)	&	326.7	&	   (	-85.2	)	&	2248	&	   (	-104	)	&	1.104	&	   (	0.031	)	&	307.0	&	   (	-55.1	)	&	2419	&	   (	-265	)	\\	
	&	CCSD$^a$	&	1.138	&	   (	0.007	)	&	329.0	&	   (	-87.4	)	&	2212	&	   (	-68	)	&	1.123	&	   (	0.013	)	&	334.0	&	   (	-82.1	)	&	2264	&	   (	-110	)	\\	
	&	CCSD(T)$^a$	&	1.145	&				&	241.5	&				&	2144	&				&	1.136	&				&	251.9	&				&	2154	&				\\	
	&	pCCD	&	1.117	&	   (	0.027	)	&	224.8	&	   (	16.7	)	&	2295	&	   (	-151	)	&	1.116	&	   (	0.020	)	&	237.9	&	   (	14.1	)	&	2317	&	   (	-163	)	\\	
	&	CCD$^b$	&	1.132	&	   (	0.013	)	&	262.8	&	   (	-21.3	)	&	2246	&	   (	-102	)	&	1.125	&	   (	0.011	)	&	288.8	&	   (	-36.9	)	&	2257	&	   (	-103	)	\\	
	&	CCSD$^b$	&	1.137	&	   (	0.007	)	&	264.0	&	   (	-22.4	)	&	2219	&	   (	-75	)	&	1.125	&	   (	0.011	)	&	289.7	&	   (	-37.8	)	&	2238	&	   (	-84	)	\\	
	&	fpLCCD	&	1.134	&	   (	0.011	)	&	229.6	&	   (	11.9	)	&	2171	&	   (	-26	)	&	1.129	&	   (	0.006	)	&	244.3	&	   (	7.6	)	&	2218	&	   (	-64	)	\\	
	&	fpLCCSD	&	1.143	&	   (	0.001	)	&	236.2	&	   (	5.3	)	&	2105	&	   (	39	)	&	1.130	&	   (	0.006	)	&	253.4	&	   (	-1.5	)	&	2155	&	   (	-2	)	\\	
	&	fpCCD	&	1.133	&	   (	0.011	)	&	235.1	&	   (	6.5	)	&	2178	&	   (	-33	)	&	1.128	&	   (	0.008	)	&	254.3	&	   (	-2.4	)	&	2228	&	   (	-74	)	\\	
	&	fpCCSD	&	1.137	&	   (	0.007	)	&	240.3	&	   (	1.2	)	&	2178	&	   (	-34	)	&	1.128	&	   (	0.008	)	&	260.7	&	   (	-8.8	)	&	2218	&	   (	-64	)	\\	
	&	exp.\cite{NIST, carbon_monoxide, irikura2007} 	&	1.128	&	   (	0.016	)	&	255.8	&	   (	-14.3	)	&	2170	&	   (	-26	)	&	1.128	&	   (	0.007	)	&	255.8	&	   (	-3.9	)	&	2170	&	   (	-16	)	\\	\hline
\end{tabular}
\end{tiny}
\end{table}

\subsection{Diatomic molecules}

As a first test case, we selected seven diatomic molecules containing only main-group elements (namely B, C, N, O, and F), which feature complex electronic structures driven by quasi-degenerate 2p orbitals. 
The dissociation process of these main-group dimers highlights the disparate interplay of nondynamic/static and dynamic electron correlation.
Specifically, the dissociation of the fluorine dimer, despite its single bond, cannot be reliably modeled with the gold standard of quantum chemistry, that is CCSD(T), as it produces an unphysical shape of the potential energy surface (PES).\cite{li1998,bytautas2007,evangelista2007,bytautas2009,geminals_lcc_2015}
A similar outcome is observed for the CCSD model in the case of the nitrogen dimer.
This molecule is known as one of the most challenging (diatomic) systems due to the triple bonding mechanism, which requires higher excitation operators in the theoretical model (as well as high angular momenta in the basis set) to reach spectroscopic accuracy.~\cite{deegan1994,li2001,li2008,wilson2011,csontos2013,seniority-cc-2014,bytautas2015,pccd-PTX}
The PESs of the carbon dimer, cyano cation, and boron nitride feature electron configurations of energetic proximity and avoided crossing.
\cite{martin1992,peterson1995,wulfov1996,peterson1997,abrams2004,sherril2005,shi2011,booth2011,wilson2011,wouters2014,pccd-2014-jcp,geminals_lcc_2015,sharma2015,gulania2018}
We also included the carbon monoxide molecule and the boron monoxide cation in our test set, which are less affected by electron correlation effects.\cite{peterson1995,geminals_lcc_2015}
The spectroscopic constants obtained for our diatomic test set by various coupled-cluster models (and basis sets) are summarized in Tables \ref{tab:3-1} and \ref{tab:3-2}.

For the F$_2$ molecule, coupled cluster methods with an RHF reference result in noticeably deeper potential energy wells.
Specifically, CCD(RHF) considerably overestimates the dissociation energy D$_{\rm e}$ with an error of 44.5 kcal/mol.
The inclusion of single excitations only slightly improves D$_{\rm e}$.
Changing the reference determinant (and hence the molecular orbital basis) to the pCCD-optimized solution results in similar shapes of the PES as obtained in conventional CCSD calculations.
Thus, switching to the pCCD reference determinant allows us to obtain CCSD accuracy by only solving the CCD equations.
The DMRG-tailored and pCCD-tailored CCD and CCSD flavours substantial improve the accuracy of the predicted potential energy well depths and vibrational frequencies.
While single-reference CCSD overestimates the potential energy well depth by 28--40 kcal/mol, this error drops down to 8--17 kcal/mol if the ansatz is tailored by multireference wave functions.
Similar gain is observed for vibrational frequencies, where tailoring decreases the difference with respect to MRCI results below the basis set error threshold.
The equilibrium bond lengths are estimated with the highest accuracy by the DMRG-tCCSD method in the pCCD-optimized orbital basis and differ from reference MRCI results by 0.007--0.015 \AA{}, which is less than the difference between the MRCI result and the experimental value.
DMRG-tCCSD in the RHF orbital basis provides similar results for correlation-consistent basis sets (cc-pVDZ, cc-pVTZ) but the errors increase when augmented basis sets (aug-cc-pVDZ, aug-cc-pVTZ) are used.

For the singlet ground-state N$_2$ molecule, all tested coupled cluster flavours achieve good accuracy in the near-equilibrium region and predict accurate bond lengths and vibrational frequencies.
However, most coupled-cluster methods fail to accurately describe the region with a stretched N--N bond and the vicinity of dissociation.
This also holds for pCCD-tailored coupled cluster methods, where the linearized models feature particularly large divergencies in the dissociation limit.
This problem is cured by both DMRG-tCCSD models in the RHF and pCCD orbital basis, which are able to (indirectly) include triple excitations in the active space spanned by the 2p orbitals.

\begin{figure}[tpb]
    \includegraphics[width=\textwidth]{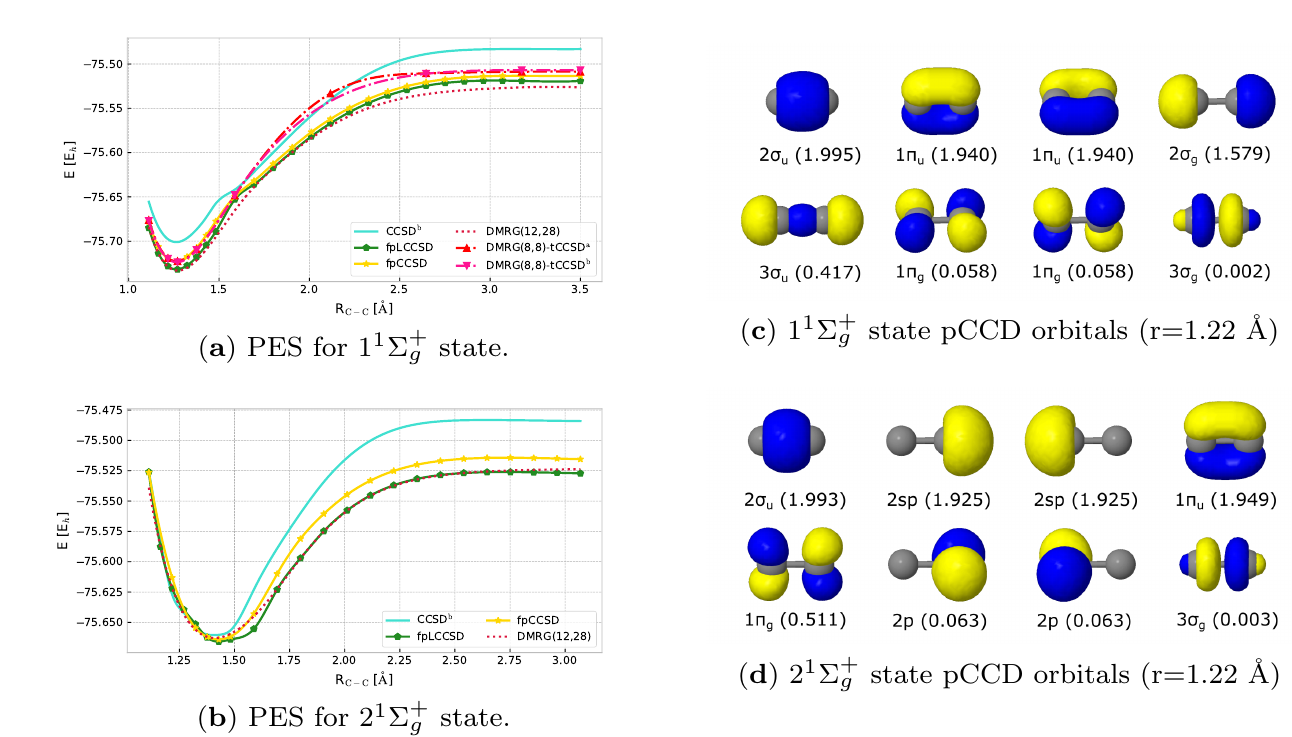} 
    \caption{The potential energy curves for the two singlet states of the carbon dimer using the cc-pVDZ basis set compared to DMRG\cite{wouters2014} reference data. The left upper panel corresponds to the ground state pCCD orbitals, while the left bottom panel was obtained for the excited-state pCCD orbitals. The superscript $a$ denotes that calculations have been performed in the canonical RHF orbital basis, while the superscript $b$ stands for pCCD-optimized orbitals. The right panels shows two sets of (valence) pCCD orbitals for the C$_2$ molecules with occupations numbers in parentheses.\label{fig:3}}
\end{figure}

The carbon dimer, cyano cation, boron nitride, and boron monoxide cation are isoelectronic diatomic species where a balanced description of electron correlation effects is required to obtain accurate energetics and spectroscopic properties.\cite{martin1992,peterson1995,pccd-2014-jcp,geminals_lcc_2015,gulania2018}
The challenge in modelling their PESs arises from the near-degeneracies of the valence orbitals and the energetic proximity of electronic configurations.\cite{wulfov1996,peterson1997,abrams2004,sherril2005,shi2011,booth2011,wilson2011,wouters2014,sharma2015,geminals_lcc_2015}
For example, the two lowest $^1\Sigma_g^+$ states of C$_2$ are affected by an avoided crossing since the $\ket{1\pi_u^4}$ configuration is favoured around the equilibrium bond length while the $\ket{3 \sigma_g^2 1\pi_u^2}$ configuration dominates in the ground state wave function in the dissociation limit.
The CN$^+$ and BN molecules exhibit an analogous bonding pattern to the carbon dimer~\cite{murrell1979,wulfov1996,peterson1997}, while BO$^+$ features a more single-reference character.\cite{peterson1997}

For the carbon dimer and cyano cation, we obtained two sets of pCCD-optimized orbitals corresponding to pCCD solutions dominated by either the $\ket{1\pi^4}$ or $\ket{3 \sigma^2 1\pi_u^2}$ determinant.
The latter causes symmetry breaking of the orbitals since the pCCD model does not describe two equivalent $\ket{3 \sigma_g^2 1\pi_u^2}$ determinants on an equal footing.
The adiabatic excitation energies of these two states are presented in Table \ref{tab:1},
while the PESs of the C$_2$ molecule obtained with the two different sets of pCCD-optimized orbitals are presented in Figure~\ref{fig:3}.
Only for DMRG-tCCSD, we were not able to optimize a different PES using the second set of pCCD orbitals as we obtained the same total energies for both orbital sets.
Our results are consistent with DMRG(12,28)\cite{wouters2014} and FCIQMC\cite{booth2011} reference data, but the PESs around the avoiding crossing region is not smooth in the case of fpCCD, fpCCSD, fpLCCD, and fpLCCSD.

\begin{table}[t] 
\caption{Adiabatic excitation energies [eV] between the singlet ground- and first excited state of the C$_2$ and CN$^+$ molecules. The acronym in parentheses indicates the molecular orbital basis employed in calculations.
The superscript $b$ denotes that calculations have been performed in the pCCD-optimized orbital basis.
}
\label{tab:1}
\centering
\begin{tabular}{lllll} \hline
	&	\multicolumn{2}{c}{C$_2$} &	\multicolumn{2}{c}{CN$^+$}\\
	&	cc-pVDZ	&	cc-pVTZ	&	cc-pVDZ	&	cc-pVTZ	\\\hline
CCD$^b$	&	1.064	&	0.997	&	1.567	&	1.567	\\
CCSD$^b$&	1.080	&	1.031	&	1.578	&	1.578	\\
fpLCCD	&	1.811	&	1.792	&	2.643	&	2.643	\\
fpLCCSD	&	1.819	&	1.774	&	2.669	&	2.669	\\
fpCCD	&	1.652	&	1.591	&	2.426	&	2.425	\\
fpCCSD	&	1.601	&	1.549	&	2.533	&	2.533	\\
DMRG(12,28)\cite{wouters2014}	&	1.913	&		&		&		\\\hline
\end{tabular}
\end{table}

\begin{figure}[tpb]
    \includegraphics[width=\columnwidth]{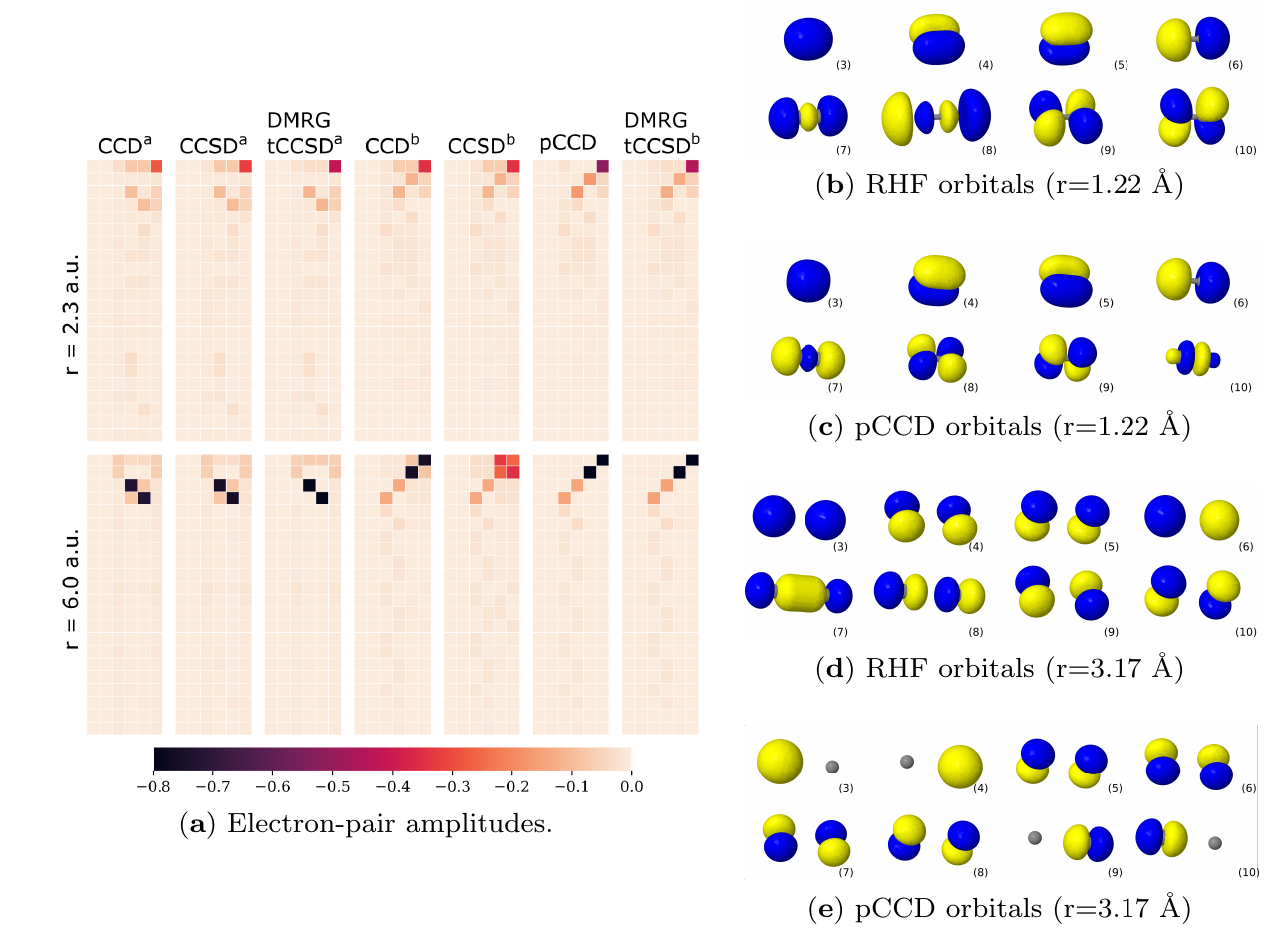}
	\caption{Electron-pair amplitudes represented as a matrix for the C$_2$ molecule obtained by CCD, CCSD, pCCD, and DMRG-tCCSD. The horizontal axis denotes occupied orbitals, while the vertical axis stands for virtual orbitals. The value of each amplitude is color-coded.
    The superscript $a$ denotes that calculations have been performed in the canonical RHF orbital basis, while the superscript $b$ indicates pCCD-optimized orbitals.}
    \label{fig:4}
\end{figure}

The electron-pair cluster (geminal) amplitudes $t_{ii}^{aa}$ of the C$_2$ molecule obtained with DMRG-tCCSD, pCCD, and (conventional) CCSD are presented in Figure~\ref{fig:4}.
The pair-amplitudes for other diatomic molecules are presented in the SI.$^\mathsection$
For the near-equilibrium geometry, the largest cluster amplitudes (in absolute value) are found for the space spanned by six 2p-type orbitals.
Two additional orbitals participate in the bond-breaking process.
In the canonical RHF basis, the electron-pair amplitudes of the conventional CCD and CCSD model agree well with the DMRG-tCCSD reference amplitudes in the equilibrium region, but disagree in the vicinity of dissociation.
For the pCCD-optimized orbital basis, the differences between pCCD and DMRG electron-pair amplitudes are small and rather of quantitative nature, while the general structure of the wave function is similar in both cases.
The CCD and CCSD electron-pair amplitudes (in both the canonical RHF and pCCD orbital basis) tend to substantially differ from the pCCD and DMRG-tailored amplitudes, which is particularly pronounced for the BO$^+$ molecule in the dissociation limit.

\begin{figure}[tb]
\centering
\includegraphics[width=0.8\textwidth]{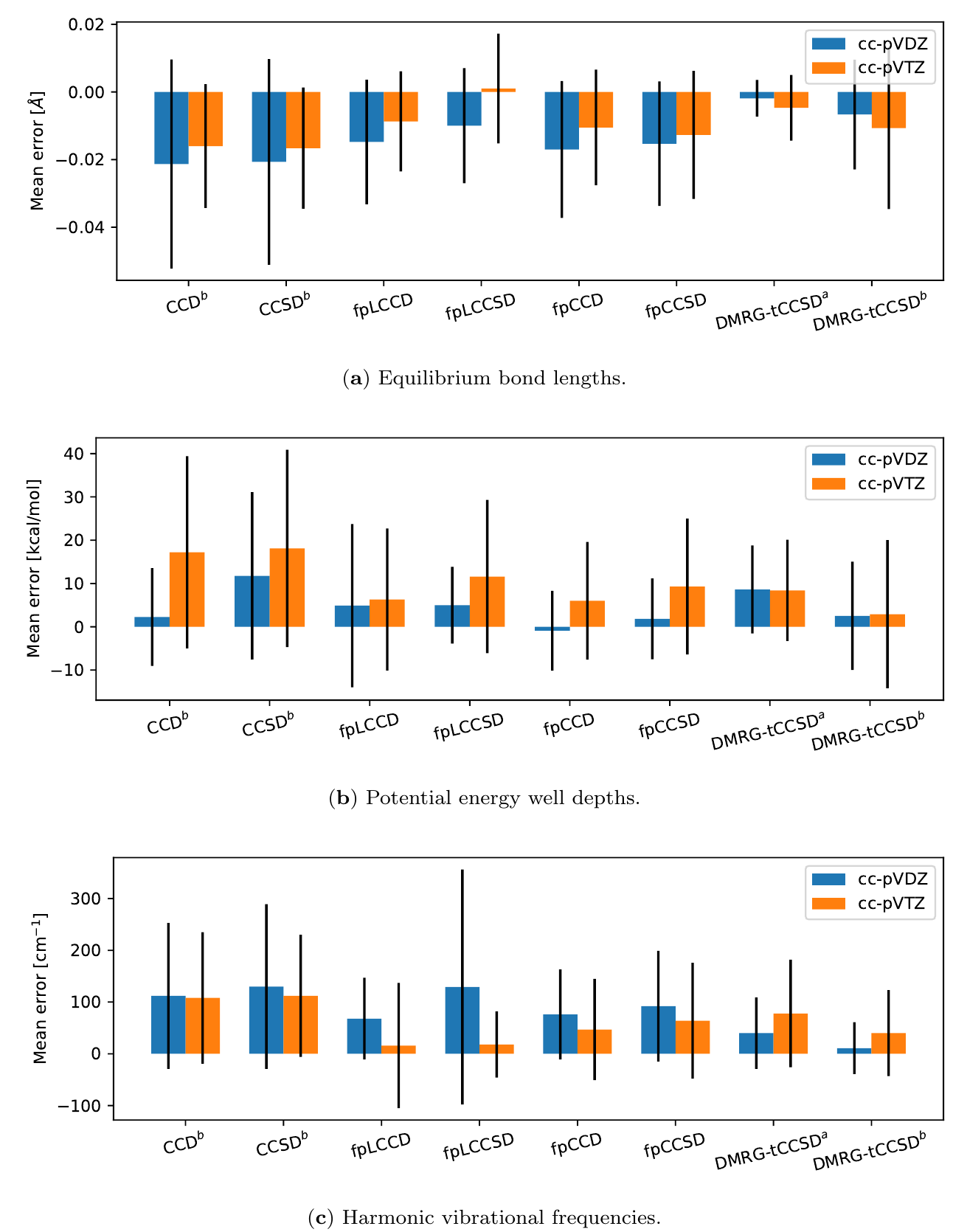}
	\caption{Mean errors with respect to accurate theoretical reference data (MRCISD+Q for BN, BO$^+$, CN$^+$, MRCI for C$_2$, CN$^+$, N$_2$, F$_2$, and CCSD(T) for CO molecule) including the standard deviation (black lines) determined for all fitted spectroscopic constants for our test set of main-group diatomics (BN, BO$^+$, C$_2$, CN$^+$, N$_2$, F$_2$, CO). The superscript $a$ denotes that calculations have been performed in the canonical RHF orbital basis, while the superscript $b$ indicates pCCD-optimized orbitals.
See also Table \ref{tab:err} for the definition of the corresponding error measures.}
\label{fig:errors}
\end{figure}

\begin{table}[t] 
\caption{Error measures determined for all fitted spectroscopic parameters (equilibrium bond lengths, potential energy well depths, and harmonic vibrational frequencies) of our test set containing main-group diatomics (BN, BO$^+$, C$_2$, CN$^+$, CO, N$_2$, F$_2$) with respect to accurate multireference methods (MRCISD+Q for BN, BO$^+$, CN$^+$, MRCI for C$_2$, CN$^+$, N$_2$, F$_2$, and CCSD(T) for the CO molecule). ME: mean error, MAE: mean absolute error, RMSD: root mean square deviation. The superscript $a$ denotes that calculations have been performed in the canonical RHF orbital basis, while the superscript $b$ indicates pCCD-optimized orbitals.}
\label{tab:err}
\begin{scriptsize}
\begin{tabular}{llrrrrrrrrr} \hline
&		&	\multicolumn{3}{c}{		$\delta$r$_\mathrm{e}$ [\r{A}]		}	&	\multicolumn{3}{c}{		$\delta$D$_\mathrm{e}$ [$\rm \frac{kcal}{mol}$]		}	&	\multicolumn{3}{c}{		$\delta\omega_\mathrm{e}$ [cm$^{-1}$]		}	\\
	&		&	ME	&	MAE	&	RMSD	&	ME	&	MAE	&	RMSD	&	ME	&	MAE	&	RMSD	\\	\hline
\parbox[t]{2mm}{\multirow{8}{*}{\rotatebox[origin=c]{90}{cc-pVDZ}}}	&	CCD$^b$	&	-0.021	&	0.021	&	0.031	&	2.3	&	9.1	&	11.3	&	112	&	112	&	141	\\
&	CCSD$^b$	&	-0.021	&	0.021	&	0.030	&	11.8	&	14.0	&	19.3	&	130	&	130	&	159	\\
&	fpLCCD	&	-0.015	&	0.015	&	0.018	&	4.9	&	13.8	&	18.9	&	68	&	68	&	79	\\
&	fpLCCSD	&	-0.010	&	0.012	&	0.017	&	5.0	&	7.1	&	8.9	&	129	&	142	&	227	\\
&	fpCCD	&	-0.017	&	0.017	&	0.020	&	-0.9	&	7.7	&	9.2	&	76	&	76	&	87	\\
&	fpCCSD	&	-0.015	&	0.015	&	0.018	&	1.8	&	7.4	&	9.3	&	92	&	92	&	107	\\
&	DMRG-tCCSD$^a$	&	-0.002	&	0.005	&	0.005	&	8.6	&	9.0	&	10.2	&	40	&	49	&	69	\\
&	DMRG-tCCSD$^b$	&	-0.007	&	0.010	&	0.016	&	2.5	&	11.2	&	12.5	&	11	&	40	&	50	\\	\hline
\parbox[t]{2mm}{\multirow{8}{*}{\rotatebox[origin=c]{90}{cc-pVTZ}}}	&	CCD$^b$	&	-0.016	&	0.016	&	0.018	&	17.2	&	18.2	&	22.2	&	108	&	108	&	127	\\
&	CCSD$^b$	&	-0.017	&	0.017	&	0.018	&	18.1	&	18.7	&	22.8	&	112	&	112	&	118	\\
&	fpLCCD	&	-0.009	&	0.012	&	0.015	&	6.3	&	13.3	&	16.4	&	16	&	98	&	121	\\
&	fpLCCSD	&	0.001	&	0.014	&	0.016	&	11.6	&	11.8	&	17.7	&	18	&	51	&	64	\\
&	fpCCD	&	-0.011	&	0.014	&	0.017	&	6	&	9.8	&	13.6	&	47	&	87	&	98	\\
&	fpCCSD	&	-0.013	&	0.016	&	0.019	&	9.3	&	13.2	&	15.7	&	64	&	95	&	112	\\
&	DMRG-tCCSD$^a$	&	-0.005	&	0.008	&	0.010	&	8.4	&	11.3	&	11.7	&	78	&	82	&	104	\\
&	DMRG-tCCSD$^b$	&	-0.011	&	0.017	&	0.024	&	2.9	&	13.8	&	17.1	&	40	&	68	&	83	\\\hline
 \end{tabular}
 {\raggedright \\
 ME (mean error) $= \frac{1}{N} \sum_i^N (x_i^{\mathrm{method}} - x_i^{\mathrm{reference}}) $ \\
 MAE (mean absolute error) $= \frac{1}{N} \sum_i^N |x_i^{\mathrm{method}} - x_i^{\mathrm{reference}}| $ \\
 RMSD (root mean square deviation) $= \sqrt{\frac{1}{N} \sum_i^N (x_i^{\mathrm{method}} - x_i^{\mathrm{reference}})^2} $ 
 \par}
\end{scriptsize}
\end{table}

\begin{figure}[tb]
\centering
	\includegraphics[width=\columnwidth]{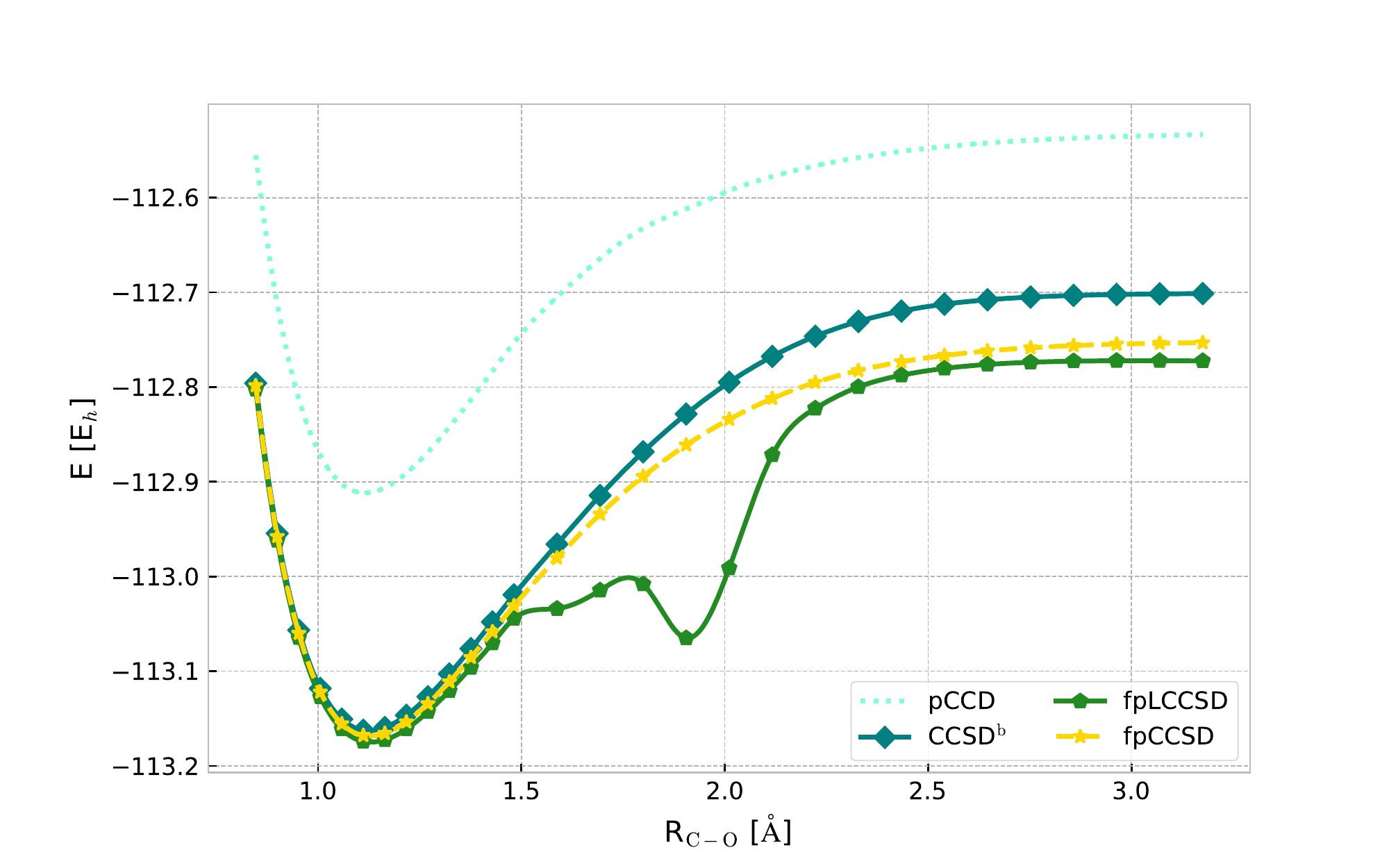}
	\caption{The potential energy curves for the CO molecule (cc-pVTZ basis set). The superscript $b$ indicates pCCD-optimized orbitals.}
	\label{fig:6}
\end{figure}

Figure~\ref{fig:6} summarizes the dissociation path of the CO molecule obtained by various CC methods.
Although the reference wave functions (RHF and pCCD) provide smooth Morse potential-shaped plots, fpLCCSD diverges in the region between 1.5--2.3 \r{A}, while CCSD in the RHF orbital basis fails in predicting a smooth PES.
Inspecting the nature and composition of the pCCD wave function, we observe that the pCCD solution becomes multireference for those bond lengths.
When the C--O distance reaches 1.48 \r{A}, the geminal coefficients for two dominant excited determinants are about $-0.1$.
They gradually increase in absolute value when the molecule is stretched reaching values of $-0.95$ and $-0.8$ for a distance of 3.17 \r{A}.
The failure of fpLCCSD may be further attributed to the linearized nature of the coupled-cluster amplitude equations, which may feature divergencies and poles in their solutions.~\cite{geminals_lcc_2017}

Figure~\ref{fig:errors} and Table~\ref{tab:err} show the mean errors including the standard deviation of the fitted spectroscopic constants (equilibrium distance, dissociation energy, and harmonic vibrational frequencies) for our test set of main-group diatomic molecules with respect to reference data.
All studied coupled cluster methods, including CCD and CCSD with pCCD-optimized orbitals, tailored CCSD approaches with both RHF and pCCD-optimized orbitals, fpCCD, and fpCCSD, overestimate equilibrium bond lengths.
In general, tailoring selected coupled-cluster amplitudes reduces the errors in all spectroscopic constants.
DMRG-tCC is the most accurate method in predicting dissociation energies (in the cc-pVTZ basis), but lies in between fpCC and fpLCC quality for equilibrium distances and vibrational frequencies. Furthermore, pCCD orbitals are slightly better in the dissociation region and thus allow us to predict more precise D$_e$ and $\omega$ values.
In general, DMRG-tCCSD(RHF) provides the smallest errors only for equilibrium bond lengths, while its performance strongly deteriorates for dissociation energies and harmonic vibrational frequencies.
Moreover, the addition of single excitations in the fpCCD model does not improve the accuracy compared to fpCCD.
Comparing the mean errors and mean absolute errors suggests that the error measures provided by tailored coupled cluster approaches are not as systematic as those obtained by single-reference coupled cluster theory, where equilibrium bond lengths are systematically too large and vibrational frequencies are systematically underestimated.
Finally, our statistical analysis on the bond-breaking process of selected main-group diatomics suggest that the results of the linearized pCCD-tCC models are of similar quality as its non-linear version.
However, pCCD-tailored wave functions (restricted to at most double excitations) are insufficient to accurately describe the dissociation pathway of molecules featuring a triple bond.
As expected, the fpLCCSD method may, however, show unphysical features in the PESs like divergencies or poles.
These divergencies can be cured by including non-linear terms in the CC amplitude equations resulting in the fpCCSD framework.

\subsection{The dissociation pathway of the Cr$_2$ dimer}

The chromium dimer and its dissociation process are widely used as a benchmark problem in quantum chemistry primarily because its complicated electronic structure and formal hextuple bond pose a remarkable challenge for present-day quantum chemical methods.\cite{cr2-1993, cr2_2009, dmrg-caspt2, cr2_2011, cr2_2016, vancoillie2016}
Even excitations of fourth order are insufficient to accurately capture electron correlation effects within a single-reference framework.\cite{dmrg2015} 
Multi-reference methods represent the most robust and trustworthy approach to study the electronic structure of Cr$_2$.
For instance, Veis \textit{et al.} report that DMRG-tCCSD effectively describes the Cr$_2$ energy around the equilibrium geometry and outperforms the conventional CCSDTQ method in terms of total energies.\cite{dmrg-tcc-2016}

\begin{figure}[tb]
\centering
	\includegraphics[width=\columnwidth]{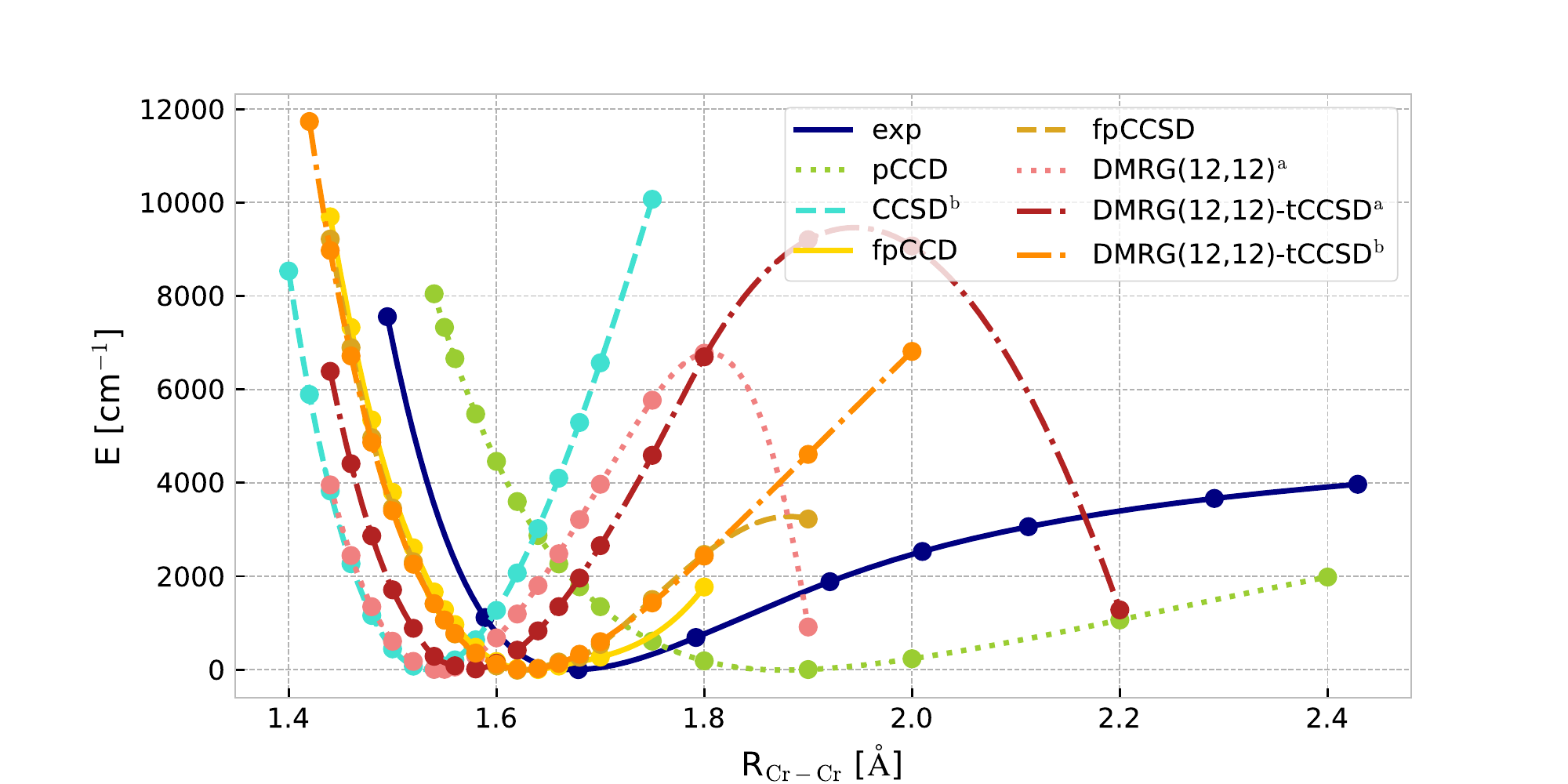}
	\caption{The potential energy curves of Cr$_2$ (cc-pVTZ basis set). The superscript $a$ denotes that calculations have been performed in the canonical RHF orbital basis, while the superscript $b$ stands for pCCD orbitals. Note that DMRG(12,12)$^{\rm b}$ yields an unbound PES and is hence not shown here.}
	\label{fig:7}
\end{figure}

Figure~\ref{fig:7} summarizes the PESs obtained with different flavours of conventional and tailored coupled-cluster theory.
pCCD provides a smooth curve, albeit overestimating the equilibrium bond length (r$_e=1.881$ \AA).
Note that pCCD does not converge for bond lengths r$_{\rm Cr-Cr} > 2.5$ \AA{}.
CCD$^b$ and CCSD$^b$ yield potential energy curves with too short bond lengths (r$_e=1.533$ \AA) and too large slopes.
fpCCD reproduces the proper shape of the PES around the equilibrium and predicts a bond length of 1.641 \AA{}, which is closest to the experimental value of 1.6788 \AA.
However, pCCD, and thus, pCCD-tailored CC theory fails in the description of the dissociation path and dissociation limit.
We should note that we encountered convergence problems for stretched Cr--Cr distances in all frozen-pair coupled-cluster calculations, while the linearized pCCD-tailored coupled cluster flavours (both fpLCCD and fpLCCSD) fail due to divergencies near the equilibrium.
Thus, those points are not shown in Figure~\ref{fig:7}.
DMRG(12,12) calculations in the RHF orbital basis do not yield a bonded PES and diverge for r$_{\rm Cr-Cr} > 1.8$ \AA{}, while DMRG(12,12) in the pCCD orbital basis provides an unphysical PES similar to CASSCF(12,12).\cite{cr2_2011, cr2_2016, vancoillie2016}
Note that the poor performance of minimal active space calculations for Cr$_2$, that is 12 electrons in 12 orbitals, is a well known problem in computational chemistry~\cite{cr2_2009, dmrg-caspt2, cr2_2016, cr2_2011, vancoillie2016}.
All investigated DMRG-tailored coupled cluster methods perform well in the near-equilibrium region.
However, for $r_{\rm Cr-Cr} > 1.9$ \AA{}, we observe overcorrelation in the RHF orbital basis, while for pCCD-optimized orbital the DMRG-tCCSD equations do not converge.
Thus, DMRG-tCCSD exploiting the minimal active space in DMRG calculations cannot be used to model the dissociation of the sextuple bond of the chromium dimer.
Further investigations are required to determine whether extending the active space can cure the problems related to overcorrelation, divergencies, and convergence difficulties in the CC amplitude equations.

\begin{table}
\caption{Computed spectroscopic constants of the chromium dimer for various coupled cluster methods and atomic basis sets.
The superscript $a$ denotes that calculations have been performed in the canonical RHF orbital basis, while the superscript $b$ indicates pCCD-optimized orbitals.}
\begin{tabular}{llll llll}
	&	r$_\mathrm{e}$ [\r{A}]		&	$\omega_\mathrm{e}$ [cm$^{-1}$]		&	r$_\mathrm{e}$ [\r{A}]		&	$\omega_\mathrm{e}$ [cm$^{-1}$]		\\	\hline
	&	\multicolumn{2}{l}{cc-pVDZ}					&	\multicolumn{2}{l}{cc-pVTZ}					\\	\hline
pCCD	&	1.884	&	264		&	1.881		&	244		\\	
CCSD$^b$&	1.542		&	950		&	1.535		&	943		\\	
fpCCD	&	1.676		&	397		&	1.637		&	425		\\	
fpCCSD	&	1.651		&	525		&	1.623		&	567		\\	
DMRG(12,12)-tCCSD$^a$	&	1.584   &	785	&	1.576	&	790	\\	
DMRG(12,12)-tCCSD$^b$	&	1.626   &	635	&	1.627	&	602	\\	
exp.	&	1.6788	\cite{bondybey1983}	&	481	\cite{casey1993}	&	1.679	\cite{bondybey1983}	&	481	\cite{casey1993}	\\	\hline
\end{tabular}
\end{table}

\subsection{Umbrella inversion of ammonia}

Theoretical models have been struggling for many years to obtain spectroscopic accuracy for the six vibrational modes of the NH$_3$ molecule.
The main reason for this struggle is an inversion mode that is characterized by a high amplitude but a low frequency.
Since this system does not feature strong electron correlation and the energy converges fast with respect to the order of the cluster operator, single-reference coupled cluster methods are sufficient to approach spectroscopic accuracy.\cite{pastorczak2015}
To assess the performance of various tailored coupled cluster flavours, we modeled the path of the conversion of the ground-state equilibrium pyramidal-shaped molecule to the planar complex.
Furthermore, the accuracy of tCC theory is benchmarked against theoretical results rather than experimental ones due to the non-monotonic behavior concerning the basis set for structural and spectroscopic properties.\cite{pesonen2001}

\begin{figure}[tb]
\centering
	\includegraphics[width=\columnwidth]{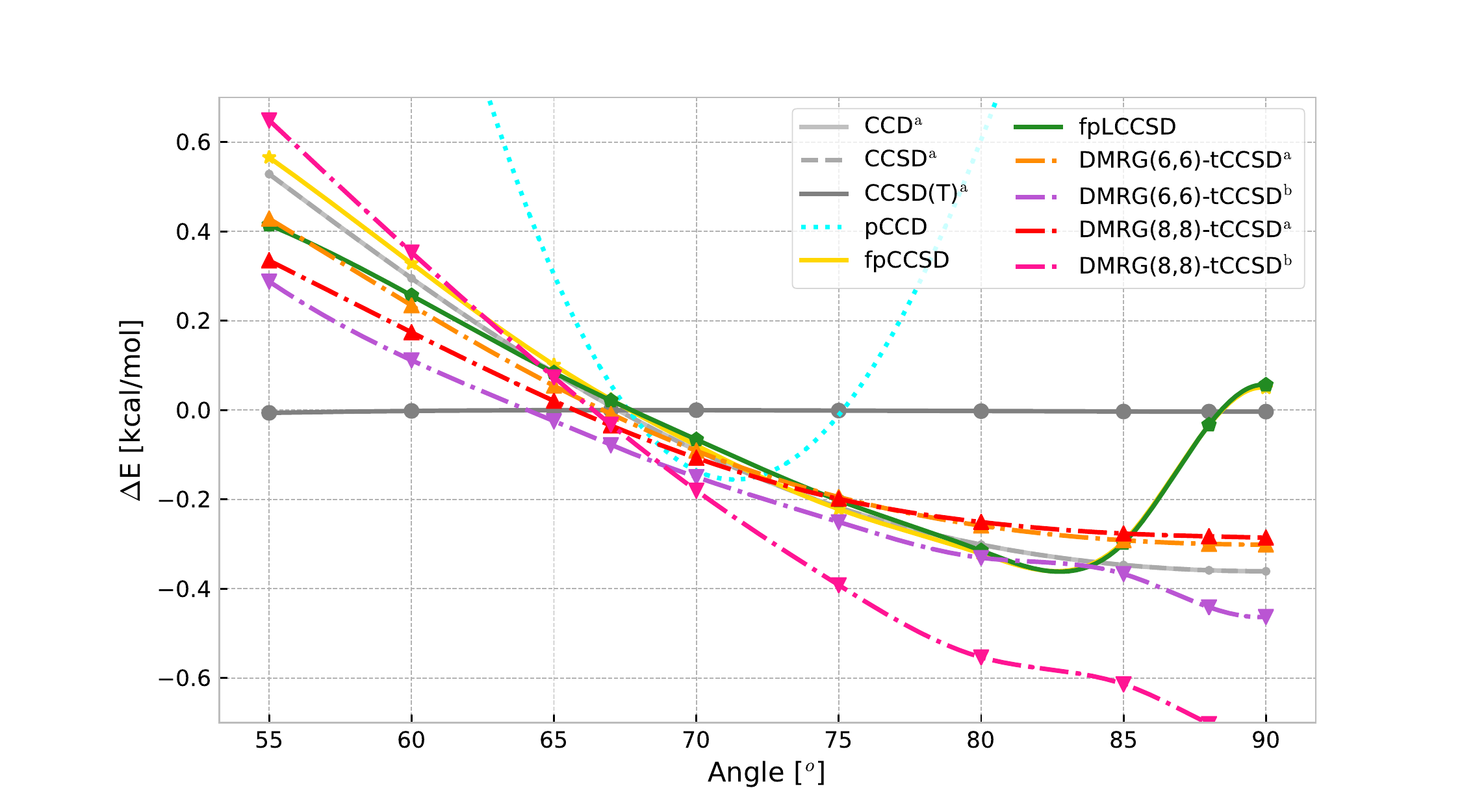}
	\caption{Ammonia umbrella inversion -- energy difference compared to CCSDT data (cc-pVTZ basis set). The type of orbitals used in DMRG calculations is denoted by upper index with $a$ standing for RHF orbitals and $b$ standing for pCCD orbitals.}
	\label{fig:ammonia}
\end{figure}

Figure~\ref{fig:ammonia} shows the potential energy curves for the umbrella inversion of the NH$_3$ complex, while Table \ref{tab:6} presents the equilibrium angles, barrier hights, and non-parallelity errors.
Both DMRG featuring a small active space and orbital-optimized pCCD (that is models capturing mostly static/nondynamic electron correlation) greatly overestimate the barrier height of the umbrella inversion process. 
Specifically, the small DMRG active space comprises six valence electrons distributed in six hybridized bonding and antibonding sp$^3$ orbitals.
We also studied a CAS that was extended by two additional sp$^3$ orbitals that feature significant values of the single orbital entropy and orbital-pair mutual information.
This overestimation originates from the fact that pCCD, DMRG(6,6), and DMRG(8,8) are insufficient to capture dynamical electron correlation effects.
Note that the shape of the PES obtained by DMRG(8,8)$^{\mathrm{b}}$ at $\alpha=90^\circ$ features an unphysical shape and diverges.
Thus, the corresponding value for D$_e$ is not shown in Table \ref{tab:6}.
For DMRG-tCCSD, we observe that augmenting the active space deteriorates the accuracy of the results compared to CCSDT reference energies.
Nonetheless, a large fraction of the missing dynamical correlation energy can be recovered by various tailored coupled cluster flavours.
Most importantly, the differences in the shape of the PESs, the equilibrium angles, and the barrier height are small with respect to CCSDT reference data for all investigated tailored CC approaches and lie within chemical accuracy as long as dynamical correlation effects have been accounted for in the theoretical model.

\begin{table*}
\caption{Equilibrium angles and barrier heights for the umbrella inversion of NH$_3$. The difference with respect to CCSDT results is given in parentheses.
The superscript $a$ denotes that calculations have been performed in the canonical RHF orbital basis, while the superscript $b$ indicates pCCD-optimized orbitals.}
\label{tab:6}
\begin{scriptsize}
\begin{tabular}{lrrrrlrrrrl} \hline
	&	\multicolumn{5}{c}{	cc-pVDZ	}		&	\multicolumn{5}{c}{cc-pVTZ	}	\\	
	&	\multicolumn{2}{c}{	$\alpha_{\rm e}$}	&	\multicolumn{2}{c}{	D$_e$ [$\rm \frac{kcal}{mol}$]}	&	NPE$^a$ [$\rm \frac{kcal}{mol}$]	&	\multicolumn{2}{c}{	$\alpha_{\rm e}$}	&	\multicolumn{2}{c}{	D$_e$ [$\rm \frac{kcal}{mol}$]}	&	NPE$^a$ [$\rm \frac{kcal}{mol}$]	\\	\hline
RHF	&	67.1	&	(	1.1	)	&	7.9	&	(	-0.7	)	&	1.9	&	68.1	&	(	1.0	)	&	6.4	&	(	-0.4	)	&	1.7	\\	
CCD	&	66.1	&	(	0.2	)	&	8.4	&	(	-0.2	)	&	0.4	&	67.4	&	(	0.4	)	&	6.5	&	(	-0.4	)	&	0.9	\\	
CCSD	&	66.1	&	(	0.2	)	&	8.4	&	(	-0.2	)	&	0.4	&	67.4	&	(	0.4	)	&	6.5	&	(	-0.4	)	&	0.9	\\	
CCSD(T)	&	65.9	&	(	0.0	)	&	8.6	&	(	0.0	)	&	0.0	&	67.1	&	(	0.0	)	&	6.9	&	(	0.0	)	&	0.0	\\	
CCSDT	&	65.9	&	(	0.0	)	&	8.6	&	(	0.0	)	&		&	67.1	&	(	0.0	)	&	6.9	&	(	0.0	)	&		\\	
pCCD	&	66.7	&	(	0.8	)	&	10.0	&	(	1.4	)	&	1.6	&	67.9	&	(	0.8	)	&	8.9	&	(	2.0	)	&	2.1	\\	
fpCCD	&	66.5	&	(	0.6	)	&	8.6	&	(	0.0	)	&	0.5	&	67.6	&	(	0.5	)	&	6.6	&	(	-0.3	)	&	1.3	\\	
fpCCSD	&	66.3	&	(	0.3	)	&	8.8	&	(	0.2	)	&	0.6	&	67.5	&	(	0.4	)	&	6.9	&	(	0.0	)	&	0.9	\\	
fpLCCD	&	66.3	&	(	0.4	)	&	8.6	&	(	0.0	)	&	0.7	&	67.6	&	(	0.5	)	&	6.5	&	(	-0.3	)	&	1.2	\\	
fpLCCSD	&	66.2	&	(	0.3	)	&	8.8	&	(	0.2	)	&	0.5	&	67.4	&	(	0.3	)	&	6.9	&	(	0.1	)	&	0.7	\\	
DMRG(6,6)$^{\mathrm{a}}$	&	68.1	&	(	2.2	)	&	8.7	&	(	0.1	)	&	2.2	&	66.8	&	(	-0.3	)	&	6.8	&	(	0.0	)	&	1.1	\\	
DMRG(6,6)-tCCSD$^{\mathrm{a}}$	&	67.2	&	(	1.3	)	&	8.4	&	(	-0.2	)	&	0.3	&	67.2	&	(	0.1	)	&	6.6	&	(	-0.3	)	&	0.7	\\	
DMRG(6,6)$^{\mathrm{b}}$	&	65.8	&	(	-0.1	)	&	10.9	&	(	2.2	)	&	0.6	&	65.7	&	(	-1.4	)	&	9.0	&	(	2.2	)	&	2.3	\\	
DMRG(6,6)-tCCSD$^{\mathrm{b}}$	&	65.8	&	(	-0.1	)	&	8.5	&	(	-0.1	)	&	0.2	&	66.1	&	(	-1.0	)	&	6.4	&	(	-0.5	)	&	0.8	\\
DMRG(8,8)$^{\mathrm{a}}$	&	65.4	&	(	-0.6	)	&	9.7	&	(	1.0	)	&	1.0	&	67.2	&	(	0.1	)	&	7.2	&	(	0.4	)	&	0.6	\\	
DMRG(8,8)-tCCSD$^{\mathrm{a}}$	&	65.4	&	(	-0.5	)	&	8.7	&	(	0.0	)	&	0.0	&	67.1	&	(	0.0	)	&	6.6	&	(	-0.3	)	&	0.6	\\	
DMRG(8,8)$^{\mathrm{b}}$	&	66.5	&	(	0.5	)	&	-	&	(	-	)	&	2.5	&	68.0	&	(	0.9	)	&	-	&	(	-	)	&	2.8	\\	
DMRG(8,8)-tCCSD$^{\mathrm{b}}$	&	65.6	&	(	-0.3	)	&	8.4	&	(	-0.2	)	&	0.4	&	67.4	&	(	0.3	)	&	6.1	&	(	-0.8	)	&	1.4	\\	\hline
\end{tabular}
{\raggedright \\ $^a$ NPE (non-parallelity error) $= \max\limits_{\alpha}(|\Delta E_{\alpha}|) - \min\limits_{\alpha} (|\Delta E_{\alpha}|) $ and $\Delta E_{\alpha} = E^{\rm CC}_{\alpha}-E^{\rm CCSDT}_{\alpha}$ \par}
\end{scriptsize}
\end{table*}

\subsection{Ethylene twist}

By twisting the dihedral angle in the ethylene molecule, we can scrutinize the flexibility of various tCC models to describe varying degrees of strong and weak electron correlation effects.
The ground state molecule in its equilibrium geometry features D$_{2h}$ symmetry and its electronic wave function is dominated by a single Slater determinant.
The orbital interaction picture changes when one CH$_2$ group is rotated by ninety degrees and the molecular point group is hence reduced to D$_{2d}$ symmetry.
For this twisted geometry, the $\pi$ and $\pi^*$ orbitals become degenerate and multi-reference approaches are required to capture static/non-dynamical electron correlation effects. 

\begin{figure}[tb]
\centering
	\includegraphics[width=\columnwidth]{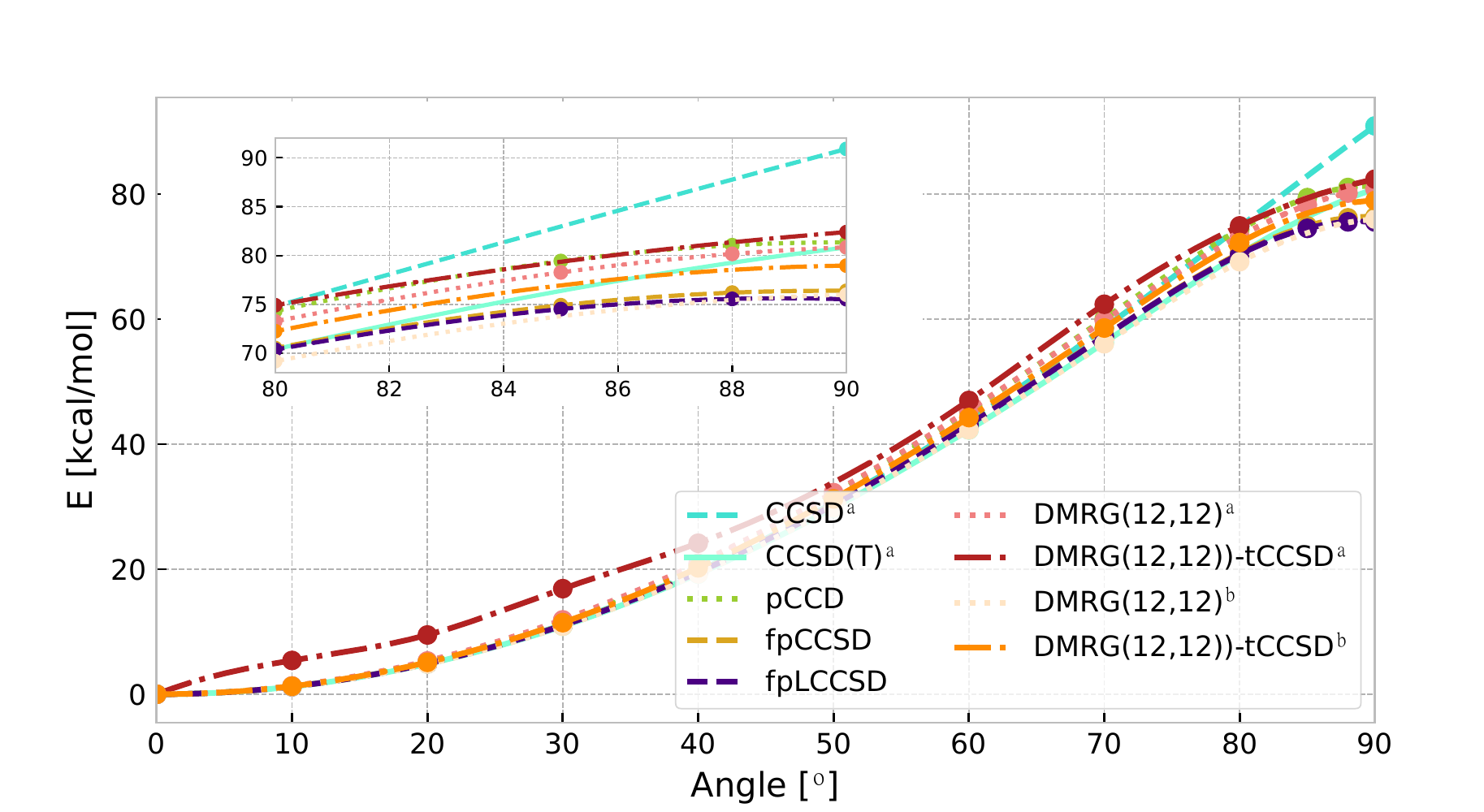}
	\caption{Potential energy surfaces for the ethylene twist (cc-pVTZ basis set) obtained by various coupled cluster methods. The superscript $a$ denotes that calculations have been performed in the canonical RHF orbital basis, while the superscript $b$ stands for pCCD orbitals.}
	\label{fig:8}
\end{figure}

Figure~\ref{fig:8} shows the potential energy curves of the ethylene torsion for various electronic structure methods.
Specifically, RHF as well as single-reference CCD and CCSD (with canonical RHF orbitals) feature an unphysical cusp in the PES for a dihedral angle of 90$^\circ$.
This cusp is a common problem encountered in the transition state when RHF orbitals are used together with some post-HF treatment that is not sufficiently accurate to account for (static/nondynamic and dynamic) electron correlation effects in a balanced way.\cite{musial2011}
A smooth potential energy curve can be obtained either with the inclusion of triple excitations, complete active space calculations containing at least two electrons and two orbitals in the active space, or with pCCD-tailored coupled cluster approaches.
Although being limited to pair excitations, the pCCD model correctly predicts that both configurations with (occupied) $\pi$ and $\pi^*$ orbitals are degenerate. 
Therefore, the orbital-optimized pCCD model is sufficient to provide a smooth reaction pathway, that is, without a cusp.
A qualitatively correct PES is also provided by DMRG(12,12) calculations for both a canonical RHF and orbital-optimized pCCD molecular orbital basis.
In general, all tCC flavours improve the shape of the PES and the barrier height for the twist without introducing unphysical effects as observed in standard single-reference approaches.

\subsection{Automerization of cyclobutadiene}

The cyclobutadiene molecule features a significant multi-reference character even in its equilibrium rectangle-shaped geometry.
The self-automerization of cyclobutadiene is a process where the carbon atoms are rearranged so that the final geometry is similar to the initial geometry, albeit rotated by 90$^\circ$.
During the self-automerization process, cyclobutadiene passes through a transition state where all C--C bonds have equal lengths and the HOMO and LUMO orbitals are exactly degenerate.
For the square geometry, the symmetry breaking of the RHF wave function affects post-HF treatments.\cite{li2009}
Specifically, single-reference coupled cluster methods tend to underestimate the weight of one of the two equivalent  $\ket{\pi^1 \pi^{*1}}$ determinants associated with the C--C double bonds.
Previous works demonstrate \cite{maksic2006, li2009, lyakh2011} that multireference approaches, like state-specific MRCCSD, outperform the so-called ``gold standard'' of quantum chemistry CCSD(T).
The selection of the reference wave function is, however, crucial since increasing the size of the active orbital space can exacerbate the quality of the results.\cite{pccd-2014-jctc}

Table \ref{tab:cbt} summarizes the automerization barrier heights obtained from various conventional and unconventional electronic structure methods.
The considerable difference in the barrier heights predicted by a perturbative treatment of triple excitations in CCSD(T) compared to full-T calculations indicates that triple excitations are important and a perturbative treatment is not sufficient.
Thus, in order to achieve reliable results using single-reference coupled cluster methods, triple excitations are to be included in the cluster operator.
Although it remains difficult to assess whether the CCSDT results are already converged with respect to the truncation order of the cluster operator as the difference between CCSD, CCSD(T), and CCSDT energies are significant, CCSDT provides a reliable description of the automerization process yielding barrier heights that agree well with experimental results and MkCCSD reference calculations.
In general, all tailored CC flavours restricted to at most double excitations overestimate the barrier height significantly, which lies up to 15 kcal/mol beyond the experimental range of 1.6--10 kcal/mol.\cite{cyclobutadiene-exp,cyclobutadiene-exp2}
The best performance is observed for DMRG-tCCSD, which predicts a barrier height between CCSD(T) and CCSDT accuracy.
Nonetheless, DMRG and DMRG-tCCSD feature a strong dependence on the choice of both the atomic basis set size and the molecular orbitals (differences amount to approximately 5--9 kcal/mol).
The method- and basis-set-dependence is smallest for (localized) pCCD-optimized orbitals (less than 1 kcal/mol).
Note that all pCCD-tCC methods provide similar barrier heights of approximately 23 kcal/mol.

\begin{table}
\caption{Barrier heights in kcal/mol for the automerization of cyclobutadiene obtained for various quantum chemistry methods and basis sets. The superscript $a$ denotes that calculations have been performed in the canonical RHF orbital basis, while the superscript $b$ stands for pCCD orbitals.}
\label{tab:cbt}
\begin{tabular}{lrr} \hline
	&	cc-pVDZ	&	aug-cc-pVDZ	\\\hline
RHF	&	28.5	&	27.2	\\
CCSD$^{\mathrm{a}}$	&	20.3	&	20.0	\\
CCSD(T)$^{\mathrm{a}}$	&	15.4	&	15.7	\\
CCSDT$^{\mathrm{a}}$	&	7.3	&	8.1	\\
pCCD	&	23.8	&	23.1	\\
fpLCCD	&	24.4	&	23.6	\\
fpLCCSD	&	24.5	&	23.8	\\
fpCCD	&	22.2	&	21.4	\\
fpCCSD	&	22.4	&	21.6	\\
DMRG(20,20)$^{\mathrm{a}}$	&	10.3	&	19.0	\\
DMRG(20,20)-tCCSD$^{\mathrm{a}}$	&	11.1	&	14.9	\\
DMRG(20,20)$^{\mathrm{b}}$ &	15.7	&	16.2	\\
DMRG(20,20)-tCCSD$^{\mathrm{b}}$	&	16.8	&	17.2	\\
NEVPT2/CAS(20,16)\cite{pccd-2014-jctc}	&	41.2	&		\\
MkCCSD\cite{Bhaskaran-Nair2008}	&	7.8	&		\\
exp.\cite{cyclobutadiene-exp}	& \multicolumn{2}{r}{	1.6 -- 10}		\\\hline
\end{tabular}
\end{table}

\subsection{Benzene distortion}

The equilibrium structure of the benzene molecule features D$_{6h}$ point group symmetry where all carbon--carbon bond lengths are equivalent.
Some approximate wave function models (including orbital-optimized pCCD \cite{pccd-2014-jctc}) tend to distort this symmetry and thus break aromaticity due to the formation of three partial double bonds caused by the localization of the three $\pi$- and $\pi^*$-orbitals.
Here, we will assess whether frozen-pair coupled cluster methods can restore the proper symmetry and hence aromaticity when imposed on top of the pCCD ansatz.
For this purpose, we scrutinize the deformation pathway of benzene as depicted in Ref. \citenum{Pierrefixe2008} where the distortion is determined by the difference between the equilibrium and distorted angle defined through some carbon atom, the center of the molecule, and the neighboring C atom.

\begin{figure}[tb]
\centering
	\includegraphics[width=0.5\columnwidth]{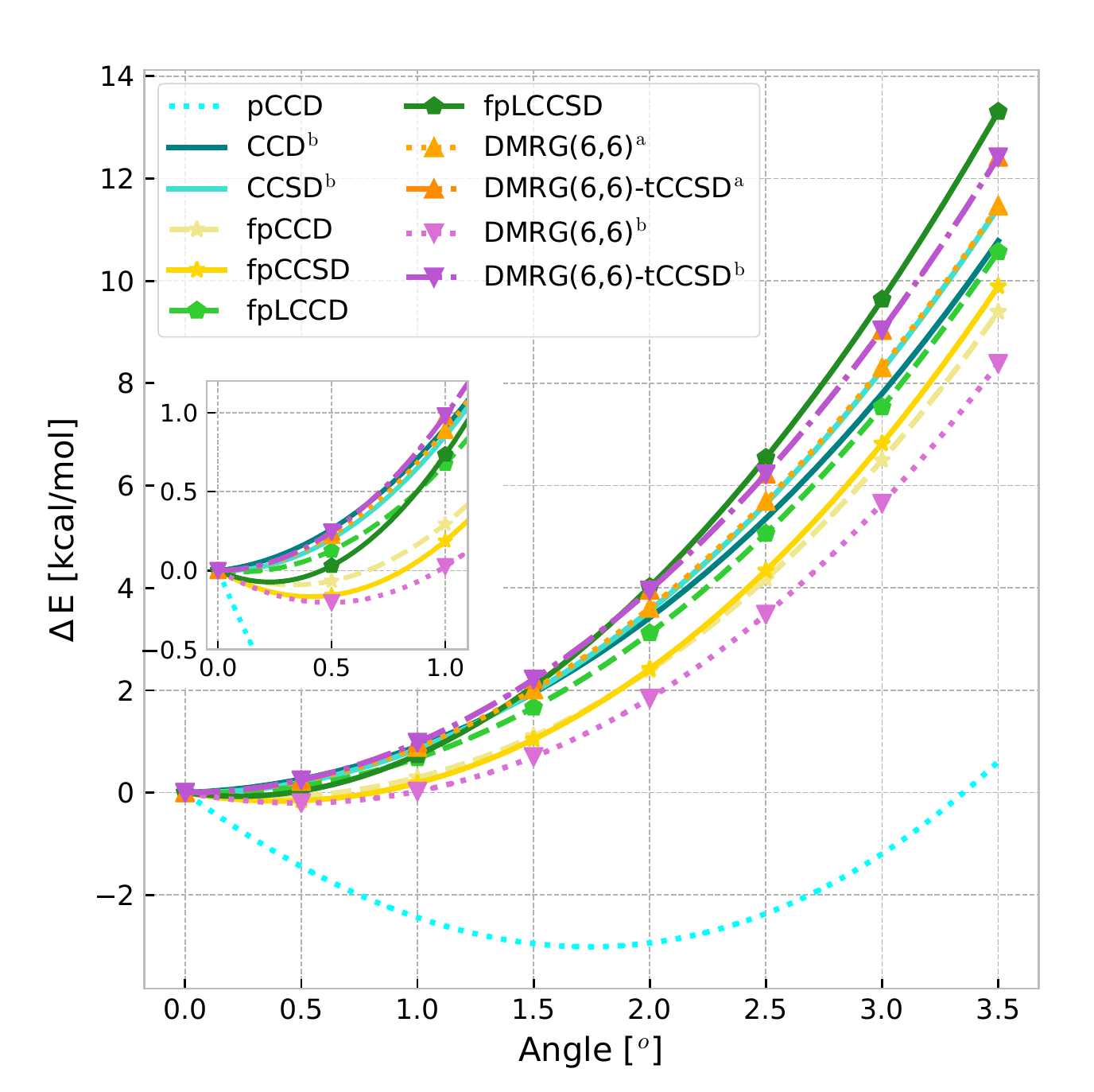}
	\caption{Potential energy surfaces for the distortion of benzene using different tailored and untailored coupled cluster methods (cc-pVTZ basis set). The superscript $a$ denotes that calculations have been performed in the canonical RHF orbital basis, while the superscript $b$ stands for pCCD orbitals.}
	\label{fig:benzene}
\end{figure}

Figure~\ref{fig:benzene} shows the PESs of the deformation process obtained from pCCD, pCCD-tLCC, pCCD-tCC, DMRG-tCC, and CCSD exploiting pCCD-optimized orbitals.
Similar to conventional CCSD (with a canonical RHF reference), which does not break the symmetry of benzene, CCSD, frozen-pair linearized CC, and DMRG-tCCSD calculations performed on pCCD-optimized orbitals correctly predict the aromatic structure (D$_{6h}$ point group symmetry) as the minimum geometry.
Although the DMRG(6,6) potential energy curves differ, we obtained almost exactly the same energies in DMRG(6,6)-tCCSD calculations exploiting both the pCCD and canonical RHF orbital basis.
On the contrary, the fpCCD and fpCCSD methods anticipate the minimum energy to be associated with some slightly deformed molecular structure, thus breaking the aromaticity of benzene.
Specifically, fpCCD and fpCCSD predict that the most stable molecular structure is distorted by approximately 0.2--0.3$^{\circ}$ and lies about 0.09--0.17 kcal/mol lower in energy than the D$_{6h}$ structure.
Most importantly, the predicted symmetry breaking is reduced compared to the pure pCCD ansatz where the distortion angle reached 1.8$^{\circ}$ and the energy difference between the aromatic and distorted structures amount to 3.01 kcal/mol.

\section{Conclusions}
\label{section:conclusions}
In this work, we scrutinized the performance of various coupled cluster methods tailored by pCCD and DMRG wave functions for various small- and medium-sized molecules (F$_2$, C$_2$, CN$^+$, BN, BO$^+$, CO, Cr$_2$, ammonia, ethylene, cyclobutadiene, and benzene) and analyzed the limitations of this computationally cheap and conceptually simple approaches.
Specifically, all conventional and DMRG-tailored CC calculations were performed for two different molecular orbital basis sets and thus reference determinants: (i) delocalized canonical RHF orbitals and (ii) localized pCCD-optimized molecular orbitals.
The active spaces in all DMRG calculations were further selected using a black-box orbital-selection protocol based on the single-orbital entropy and orbital-pair mutual information.
If pCCD-optimized orbitals are chosen as molecular orbital basis in DMRG calculations, an entropy- and/or correlation-based active space can be furthermore straightforwardly and cheaply selected from the complete set of orbitals as these measures are readily available at the end of an orbital-optimized pCCD calculation~\cite{dmrg-2016}.
This feature greatly facilitates and speeds up DMRG calculations with respect to selecting proper active orbital spaces, DMRG convergence, and computational cost (that is, the maximum number of bond dimensions $m$ required for energy convergence).

Tailored CC approaches noticeably improve the quality of spectroscopic properties, that is equilibrium bond lengths, harmonic vibrational frequencies, and dissociation energies, in comparison to their single-reference counterparts CCD and CCSD.
Although restricting the cluster operator to at most double excitations is insufficient to reach the accuracy of the more expensive CCSDT, MRCC, or MRCI methods, the mean errors in spectroscopic constants can be reduced to 0.005 \AA{} for bond lengths, 2.5 kcal/mol for dissociation energies, and 40 cm$^{-1}$ for harmonic vibrational frequencies.
Thus, tailored coupled cluster theory constitutes a promising alternative to the conventional CCSD approach if the inclusion of triple and higher excitation operators is computationally too demanding.
Furthermore, for F$_2$, C$_2$, CN$^+$, BN, BO$^+$, CO, ammonia, and ethylene, we obtained congruent coupled cluster electron-pair amplitudes and spectroscopic constants for both DMRG- and pCCD-tailored approached.
This indicates that electron-pair amplitudes are qualitatively well described by the rather simple and cheap pCCD approach.
Statistically, the performance of DMRG-tCCSD in predicting reliable spectroscopic constants lies between fpCCSD and its linearized version fpLCCSD, while fpCC(S)D and fpLCC(S)D typically provide results of similar accuracy (like relative energy differences and spectroscopic constants).
The major drawback of linearized fpCC methods is that divergencies and poles can be encountered when solving the set of the (linearized) amplitude equations.
These divergencies and poles disappear if non-linear terms are added in the amplitude equations thus resulting in fpCC-type methods.
Finally, the matrix product state ansatz optimized by the DMRG algorithm is a more flexible reference wave function for externally-corrected CC methods than pCCD as it correctly describes the dissociation of the triple bond of the nitrogen dimer and does not break in the region of avoided crossings.
The performance of DMRG-tCC is, however, dependent on the choice of the molecular orbital basis: while for some cases (like equilibrium bond lengths, automerization of cyclobutadiene, benzene distortion) delocalized canonical RHF orbitals yield results that agree better with reference data (both from experiment and multireference calculations), DMRG-tCC exploiting localized pCCD-optimized orbitals performs better for others (like vibrational frequencies, dissociation energies, Cr$_2$).
However, none of the studied tailored CC methods was able to predict the correct barrier height for the automerization of cyclobutadiene or to reliably describe the complete potential energy surface of the chromium dimer.
Most likely, full triple excitations are required to reach chemical accuracy for such challenging systems.

\begin{acknowledgement}
A.~L.~and K.~B.~thank for financial support from the National Science Centre, Poland (SONATA BIS 5 Grant No.~2015/18/E/ST4/00584).
A.~L. acknowledges funding from Interdisciplinary Doctoral School \textit{Academia Copernicana}.
M.~M. and \"{O}.~L.~acknowledges financial support from the Hungarian National Research, Development and Innovation Office (Grant Nos. K120569 and K134983), the Hungarian Quantum Technology National Excellence Program (Grant No.~2017-1.2.1-NKP-2017-00001) and the Hungarian Quantum Information National Laboratory (QNL).
\"{O}.~L.~acknowledges financial support from the Alexander von Humboldt foundation.
Calculations have been carried out using resources provided by Wroclaw Centre for Networking and Supercomputing (http://wcss.pl), Grant No.~412.
The development of the DMRG libraries was supported by the Center for Scalable and Predictive methods for Excitation and Correlated phenomena (SPEC), which is funded from the Computational Chemical Sciences Program by the U.S. Department of Energy (DOE), at Pacific Northwest National Laboratory. 
\end{acknowledgement}



\providecommand{\latin}[1]{#1}
\makeatletter
\providecommand{\doi}
  {\begingroup\let\do\@makeother\dospecials
  \catcode`\{=1 \catcode`\}=2 \doi@aux}
\providecommand{\doi@aux}[1]{\endgroup\texttt{#1}}
\makeatother
\providecommand*\mcitethebibliography{\thebibliography}
\csname @ifundefined\endcsname{endmcitethebibliography}
  {\let\endmcitethebibliography\endthebibliography}{}

\end{document}